\documentclass[12pt]{article}
\pdfoutput=1
\usepackage{jheppub}
\usepackage{amsmath,subcaption,comment}
\usepackage{mathrsfs,bbm}
\usepackage{amsfonts,booktabs,xcolor}
\usepackage{tikz}
\usetikzlibrary{decorations.markings}

\newcommand{\cF}{\mathcal F}

\author[a]{Arash Arabi Ardehali,}
\author[b]{Daniel J. Resnick$\,$}

\affiliation[a]{Department of Physics, Sharif University of Technology,\\
P.O. Box 11155-9161, Tehran, Iran}

\affiliation[b]{C.N. Yang Institute for Theoretical Physics, Stony Brook University,\\ Stony Brook, NY 11794, USA}

\emailAdd{a.a.ardehali@gmail.com}\emailAdd{daniel.j.resnick98@gmail.com}

\title{Perturbative Coulomb branches on $\mathbb{R}^3\times S^1$: the global D-term potential}

\date{}

\begin{document}

\abstract{We develop a Wilsonian description of perturbative
D-terms in 4d $\mathcal N=1$ gauge theories on
$\mathbb R^3\times S^1$ with chiral Kaluza--Klein spectra.
A closed-form, one-loop-exact coefficient of the auxiliary D fields,
including 4d Fayet--Iliopoulos parameters and globally valid throughout the holonomy space, determines the
leading holonomy potential away from massless thresholds. We establish
uniform weak-coupling control there and match across thresholds by retaining the light fields. Gauge-anomaly cancellation renders the Coulomb-slice vacuum equations affine within each chamber bounded by the thresholds, enabling exact linear analysis of their zero set. We present also a numerical reconstruction pipeline through DBSCAN clustering of approximate zeros, LPCA denoising, and RANSAC hyperplane detection. Complementing the rank-one applications of the accompanying Letter, we study flavor-twist dependence in the $\mathrm{SU}(2)$ $\mathcal N=4$ theory, and illustrate the numerical reconstruction in higher-rank R-twisted Argyres--Douglas reductions of interest in the SCFT/VOA correspondence. We also relate the same D-term data to leading stationary-phase equations of supersymmetric partition functions and a decompactification limit of the torus twisted superpotential.}

\maketitle

\section{Introduction and summary}\label{sec:intro}

Circle compactification makes the gauge holonomy a dynamical variable
and can reveal low-energy sectors invisible near trivial holonomy.
In four-dimensional $\mathcal N=1$ theories on $\mathbb R^3\times S^1$,
supersymmetry cancels the one-loop holonomy potential \cite{Gross:1980br,Poppitz:2021cxe}, but does not in general guarantee perturbative flatness.
A chiral Kaluza--Klein (KK) spectrum induces Fayet--Iliopoulos (FI)
and Chern--Simons (CS) couplings. These enter linearly in the
three-dimensional auxiliary D fields and generate a scalar potential
when the auxiliary fields are eliminated \cite{Dorey:1999rb,Tong:2000ky,Intriligator:2013lca}.

Chirality here is a property of the full KK spectrum. It can arise
from four-dimensional representations or from asymmetric flavor or
R-symmetry twists, even when the untwisted theory is vectorlike.
Conversely, charge-conjugate modes with equal absolute real masses
cancel in the D-linear coefficient. Table~\ref{tab:SUSY_chirality}
summarizes the distinction, assuming that holonomy directions are flat at tree-level and that the boundary conditions around the circle
preserve the stated supersymmetry.
\begin{table}[t]
\centering
\begin{tabular}{p{.43\linewidth}p{.21\linewidth}p{.23\linewidth}}
\toprule
4d theory on $\mathbb{R}^3\times S^1$ & Perturbative\ \ \ \ \ \ \ \ 
potential & Nonperturbative potential\\
\midrule
$\mathcal{N}\ge2$ & Absent & Absent\\
$\mathcal{N}=1$, nonchiral KK spectrum & Absent & Possible\\
$\mathcal{N}=1$, chiral KK spectrum & $\quad{\color{blue}\checkmark}$ & Possible\\
\bottomrule
\end{tabular}
\caption{Quantum lifting of a classically flat holonomy direction.
The checkmark concerns the perturbative D-terms studied here;
nonperturbative F-terms are outside our scope.}
\label{tab:SUSY_chirality}
\end{table}

Determination of the low-energy sectors is a global problem: the fields that can be integrated
out change as the holonomy crosses a massless threshold.
We solve the perturbative matching problem throughout the holonomy space by combining a
closed-form KK sum for the D-linear coefficient with the patchwise
Wilsonian construction of Ref.~\cite{Ardehali:2021irq}.
The companion Letter~\cite{CompanionPRL} presents the resulting
framework and two rank-one applications. This paper supplies detailed
derivations, control estimates,
higher-rank examples, and
the relation to partition-function asymptotics.

Our applications are complementary to those of the Letter.
Flavor-twisted $\mathrm{SU}(2)$ $\mathcal N=4$ super-Yang--Mills
theory illustrates changes in the Coulomb vacua as twist
parameters vary. The R-twisted $(A_1,A_4)$ reduction gives a
rank-two example of exact chamberwise analysis, while the
$(A_1,A_6)$ and $(A_1,A_7)$ reductions illustrate numerical reconstruction in
three independent holonomy variables. These calculations identify zeros of perturbative
potentials. Their interpretation requires distinguishing
outer Coulomb or topological sectors away from massless thresholds, from inner theories retaining
charged matter or enhanced gauge symmetry at a threshold.

Our focus is the perturbative matching and the vacuum equations
in controlled weak coupling regimes. Small 4d coupling at the compactification scale controls
the matching as well as the vacuum structure away from thresholds, but need not control 3d
infrared (IR) dynamics near thresholds. We supplement the $(A_1,A_4)$ example in
Section~\ref{subsubsec:A1A4} with known IR results to
complete the picture, as in the Letter.
Strongly coupled 3d IR dynamics is otherwise outside our
focus. Nonperturbative F-terms, including possible monopole
superpotentials, are beyond the scope of the present work.

\subsection*{Review of earlier results}

Write the holonomy as $\exp(2\pi i\boldsymbol{x})$, with
$\boldsymbol{x}$ in the Cartan algebra modulo the cocharacter lattice
and the Weyl group. In an integral lattice basis, its independent
coordinates are 1-periodic. The dimensionful 3d vector-multiplet
scalars obey

\begin{equation}
	\sigma_j=\frac{x_j}{R_1},
\end{equation}
and parametrize (absent 4d FI parameters, part of) the classical vacuum moduli space---see Sections~\ref{subsec:terms} and \ref{sec:scalar_pot}.

\begin{figure}[ht]
    \centering
    \begin{subfigure}{0.49\textwidth}
    \centering
\includegraphics[width=\textwidth]{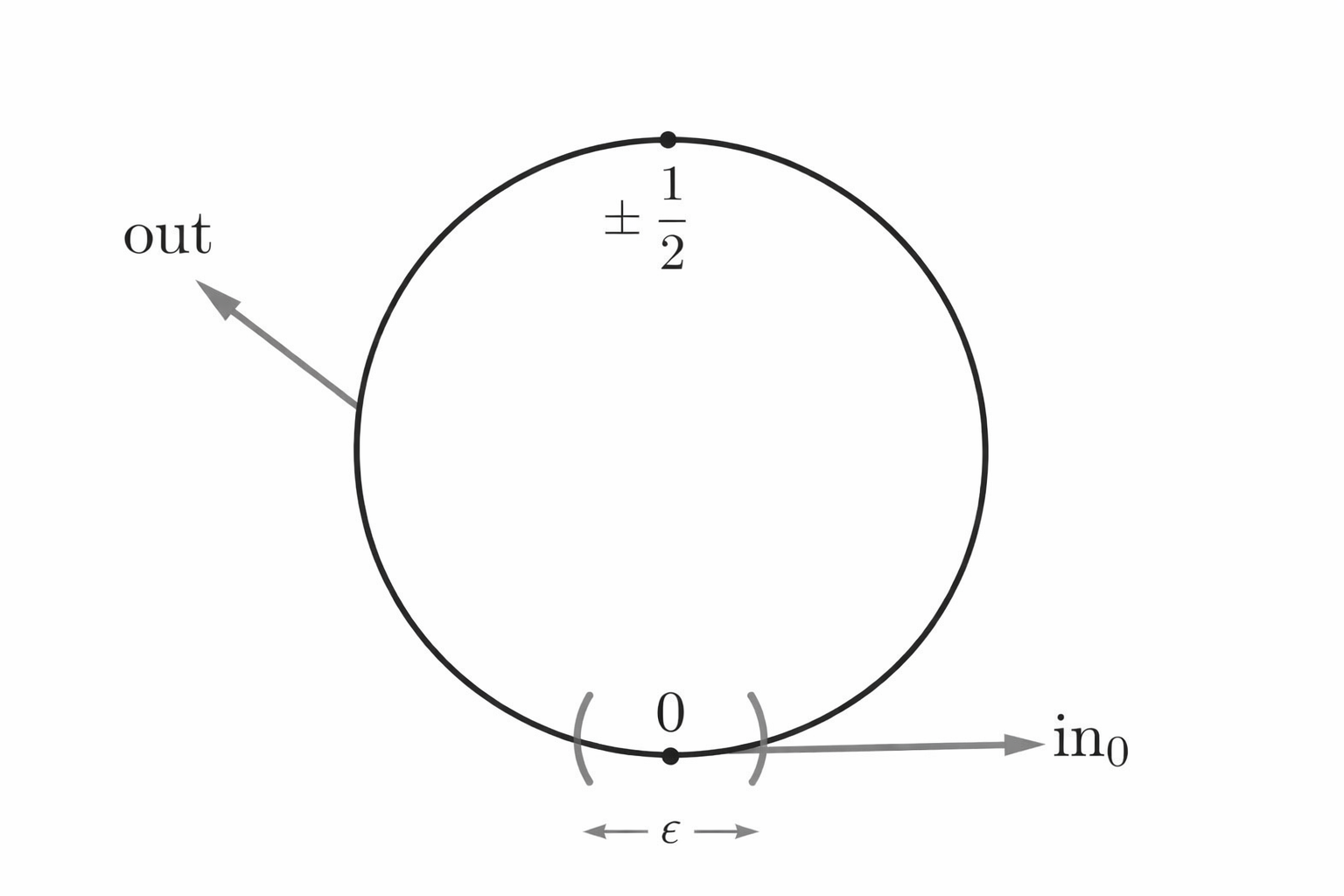}
\end{subfigure}%
\hfill
 \begin{subfigure}{0.51\textwidth}
    \centering
     \includegraphics [width=\textwidth]{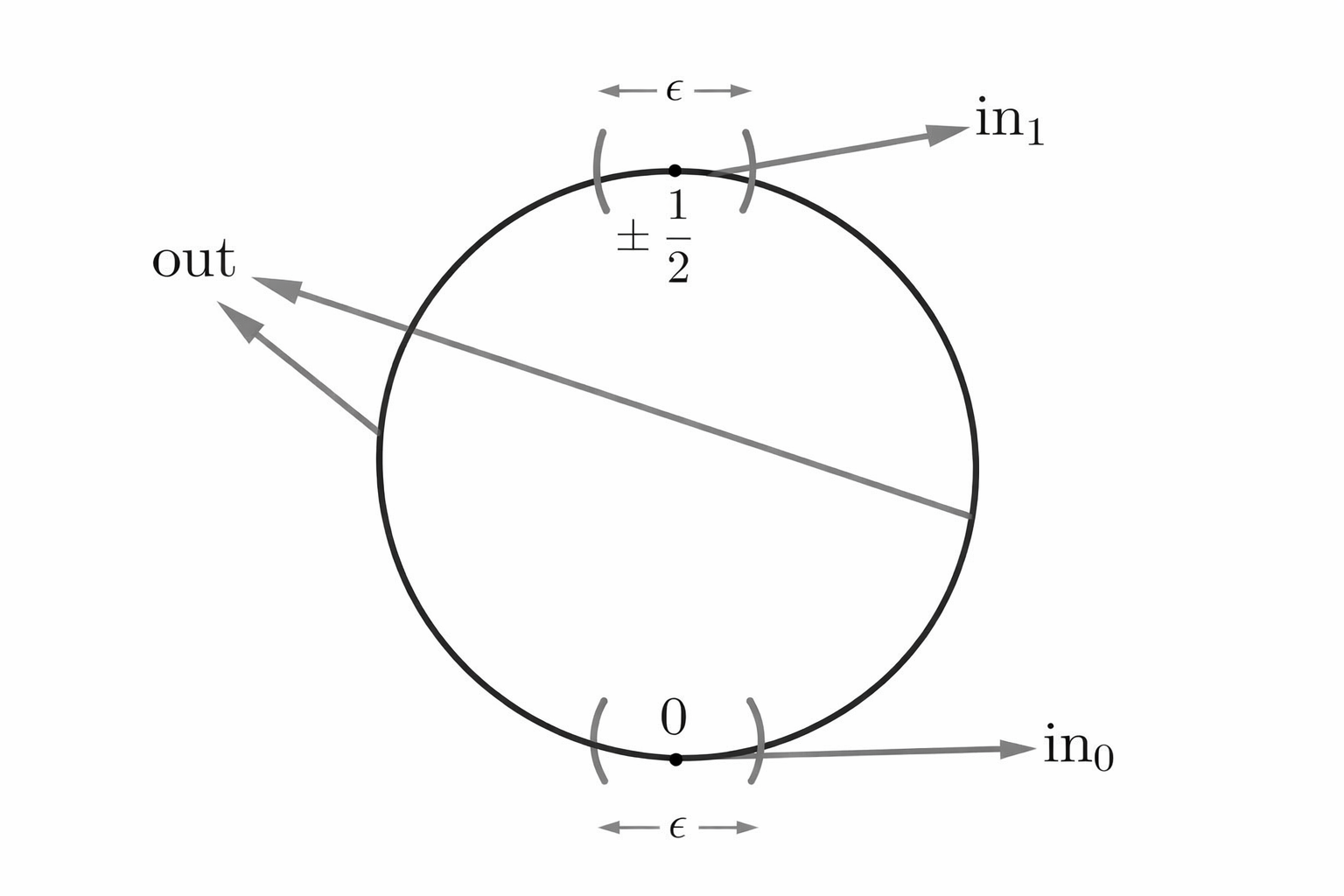}
\end{subfigure}   
    \caption{Decomposition of the classical Coulomb branch for a $U(1)$ gauge theory such as SQED (left) and an $\mathrm{SU}(2)$ gauge theory such as SQCD (right). Periodic conditions around $S^1$ are assumed for simplicity, and zero FI coupling for SQED. The light degrees of freedom on the outer patch in both cases comprise 3d $\mathcal{N}=2$ super-Maxwell. In the SQED example the inner patch in$_0$ is governed by 3d $\mathcal{N}=2$ SQED. In the SQCD example, the inner patch in$_0$ is governed by 3d $\mathcal{N}=2$ $\mathrm{SU}(2)$ SQCD, and the (would-be center-symmetry image) inner patch in$_1$ is governed by 3d $\mathcal{N}=2$ $\mathrm{SU}(2)$ SYM.}
    \label{fig:U1-SU2}
\end{figure}

As we shall review in Section~\ref{sec:scalar_pot}, in an effective field theory (EFT) approach, the moduli space of $x_j$ should be decomposed into multiple patches. The low-energy description is consequently a collection of matched 3d effective field theories, each valid on its own patch. See Figure~\ref{fig:U1-SU2} for rank-one examples. Most aspects of Figure~\ref{fig:U1-SU2} should be familiar from Coulomb branch physics; we only remark that on the inner patch in$_1$, it is W-boson modes with nonzero KK number that become light and are responsible for $U(1)\to\mathrm{SU}(2)$ gauge enhancement.

On each patch we seek the perturbative couplings of a 3d
$\mathcal N=2$ EFT. The cancellation for a paired KK spectrum
underlies familiar analyses of nonchiral reductions~\cite{Aharony:1997bx}, but in a chiral setting there is a D-term potential \cite{Dorey:1999rb,Tong:2000ky}. Consider first a rank-one scenario with a $U(1)$ EFT on the outer patch, whose 3d $\mathcal{N}=2$ vector multiplet $(A_\mu,\sigma,\lambda,D)$ contains besides the gauge field $A_\mu$ and the Coulomb branch scalar $\sigma$, the gaugino $\lambda$ and the auxiliary field $D$. Assuming an effective Chern-Simons (CS) level $k^\text{eff}$ and an effective FI-parameter $\zeta^\text{eff}$, we have the D-term effective potential
\begin{equation}
	\mathcal{L}_D=\frac{1}{2e_0^2}D^2-\frac{\zeta^\text{eff}}{4\pi}D-\frac{k^\text{eff}}{4\pi}\sigma D\xrightarrow{D=\frac{e_0^2}{4\pi}(\zeta^\text{eff}+k^\text{eff}\sigma)}V_D^{(0)}=\frac{e_0^2}{32\pi^2}(\zeta^\text{eff}+k^\text{eff}\sigma)^2,\label{eq:L_D}
\end{equation}
with $e_0$ the the EFT gauge coupling. In higher-rank scenarios, the outer-patch gauge group is $U(1)^{r_G}$, and the above expression generalizes to\footnote{We assume the gauge couplings for the different gauge factors are equal. Relaxing this to the more natural setting with comparable but not necessarily equal couplings is straightforward.}
\begin{equation}
	V_D^{(0)}=\frac{e_0^2}{32\pi^2}\sum_{i=1}^{r_G}F_i^2,\qquad F_i=\zeta^\text{eff}_i+\sum_{j=1}^{r_G}k^\text{eff}_{ij}\sigma_j,\label{eq:Vsc_intro}
\end{equation}
with $\zeta^\text{eff}_i$ the effective FI parameter of the $i$th $U(1)$ factor, and $k^\text{eff}_{ij}$ the effective gauge-gauge CS coupling between the $i$th and $j$th $U(1)$. See the original papers \cite{Dorey:1999rb,Tong:2000ky}, as well as the later elaborations by Intriligator and Seiberg \cite{Intriligator:2013lca}. In Appendix~\ref{app:R3_from_S3}, we show how this scalar potential is related to a function $Q^{(3)}(\boldsymbol{\sigma})$ arising in the large-radius limit of the $S^3$ partition function. This can be compared with Russo's discussion of how the 4d $\mathcal{N}=2$ prepotential on $\mathbb{R}^4$  arises in the large-radius limit of the $S^4$ partition function \cite{Russo:2014nka}, or with the closely related derivation of the 4d $\mathcal{N}=2$ prepotential from the $\epsilon\to0$ limit of the Nekrasov partition function \cite{Nekrasov:2002qd}.

We seek the coefficient $F_j$ throughout the holonomy space,
including the changes required when charged modes become light.
The induced CS levels were studied in, for example,
Refs.~\cite{Poppitz:2008hr,DiPietro:2016ond,Corvilain:2017luj}.
The KK sum for the D-linear coefficient and the local FI
and CS couplings near holonomy saddles were obtained in
the torus-reduction analysis of Hwang, Lee, and Yi
\cite[Sec.~3.3]{HwangLeeYi2018}.\footnote{We became aware of the results in Sec.~3.3 of \cite{HwangLeeYi2018} only after the first version of the present work appeared on arXiv. The features that most clearly distinguish our work from theirs are the inclusion of nonzero 4d FI parameters, an explicit Wilsonian cutoff, a sharp delineation of the perturbatively controlled regimes, and incorporation of charged condensates or loop corrections as controlled mechanisms for canceling the D-coefficient.}
Here we give an independent derivation on
$\mathbb R^3\times S^1$, including 4d FI inputs, and
organize the effective description using an explicit
Wilsonian matching cutoff.
We establish uniform control of the leading outer
potential in \eqref{eq:Vsc_intro}, implement matching
by subtracting the retained light modes from the KK sums,
and develop control criteria for the inner theories'
Coulomb, Higgs, and mixed sectors, building on
Ref.~\cite{Intriligator:2013lca}.

An alternative route to chiral reductions starts with a nonchiral
parent and decouples additional fields by real masses
\cite{Aharony:2013dha,Amariti:2015kha}. This is effective in the
sectors tracked by the deformation, but need not exhibit every
sector at nontrivial holonomy. A global expression for $F_j$
provides the data for finding all outer perturbative zeros and
the FI and CS couplings needed to analyze the remaining patches.

\subsection*{Overview of the new result}

The common starting point of this paper and the companion
Letter~\cite{CompanionPRL} is the closed-form outer-patch expression
\eqref{eq:Fj} for
$F_j=\zeta_j^\text{eff}+k_{ij}^\text{eff}\sigma_i$.

The calculation of $\zeta^\text{eff}_j$ is, similarly to that of $k^\text{eff}_{ij}$, by summing the 1-loop-exact contributions from heavy charged KK modes on the Coulomb branch. Each such KK mode makes a 1-loop contribution (see e.g.~\cite{Intriligator:2013lca})
\begin{equation}
    \delta\zeta^\text{1-loop}_j=\frac{1}{2}\,\rho_j\, m\ \mathrm{sign}(m+\rho\cdot\boldsymbol{\sigma}), \label{eq:FI_correction}
\end{equation}
to $\zeta^\text{eff}_j$. Here $m$ is the holonomy-independent real-mass parameter, so the full mode mass is $M=m+\rho\cdot\boldsymbol\sigma$, and $\rho_j$ is its charge under the $j$th $U(1)$ in the Cartan of the gauge group. This expression should be summed over all KK modes, and the sum should be zeta-regularized to give the full correction $\zeta^\text{1-loop}_j$.

Let the renormalized 4d FI inputs be $\zeta_j^{(4)}$. In terms of the dimensionless parameters $\tilde{\zeta}^{(4)}_j=(2\pi R_1)^2\zeta^{(4)}_j,$ the 3d EFT would then inherit tree-level FI-parameters $\zeta_j=\tilde{\zeta}^{(4)}_j/R_1$. Since the KK modes generically have $m\propto 1/R_1,$ both the correction in Eq.~\eqref{eq:FI_correction} and the tree-level 3d FI-parameters $\zeta_j$ are proportional to $1/R_1.$ In the combination $F_j=\zeta^\text{eff}_j+k^\text{eff}_{ij}\sigma_i$, also $\sigma_i=x_i/R_1$ is proportional to $1/R_1$. We thus get $F_j\propto 1/R_1$, as seen in Eq.~\eqref{eq:Fj} below. Since $e_0^2\sim g_4^2/R_1$, the leading 3d potential scales as $g_4^2/R_1^3\sim e_0^2/R_1^2$ at fixed dimensionless twists and FI inputs; see Eq.~\eqref{eq:Vsc}.

We use zeta-function regularization for the KK sums, with
renormalized FI inputs and local counterterms fixed in the same
scheme. Supersymmetric partition functions are usually computed in the same scheme, and their flat-space asymptotics are indeed compatible with our results as demonstrated in Appendix~\ref{app:index}. We also show compatibility with the $\mathbb{R}^2\times T^2$ twisted superpotential in the decompactification limit $T^2\to\mathbb{R}\times S^1$ in Appendix~\ref{app:QfromW}. For more on regularization schemes and counter-terms see \cite{Corvilain:2017luj,Poppitz:2020tto} and Appendix~\ref{app:KKren}.

Near a massless threshold, the Wilsonian description must retain
the light charged modes explicitly and subtract their would-be contributions
from the effective couplings
\cite{Ardehali:2021irq,Intriligator:2013lca}.
If $L_{\mathrm{in}}$ denotes the retained chiral modes, with
charges $\rho_i$ and real masses $M_\rho$, the matching relations
are
$F_i^{\mathrm{in}}=F_i-\frac12\sum_{\rho\in L_{\mathrm{in}}}
\rho_i|M_\rho|$ and
$k_{ij}^{\mathrm{in}}=k_{ij}^{\mathrm{out}}
-\frac12\sum_{\rho\in L_{\mathrm{in}}}
\rho_i\rho_j\operatorname{sign}(M_\rho)$.
The resulting inner couplings are smooth across the retained
thresholds; gauge-anomaly cancellation makes
$F_i^{\mathrm{in}}$ affine on each inner patch.
Although $F_i^{\mathrm{in}}$ and the continued outer coefficient
$F_i$ agree wherever all retained charged chirals are massless,
their zero sets need not agree elsewhere on the patch.
At gauge-enhancement walls, the light vector multiplets must
also be restored, and the Cartan couplings organized into
couplings of the full unbroken gauge group.
Section~\ref{subsubsec:retained} implements this matching
directly in the KK sums.

The inner-patch vacuum equations involve both these matched
couplings and the retained fields \cite{Intriligator:2013lca}. With canonical matter
kinetic terms, classical D-flatness requires
$F_A^{\mathrm{in}}=2\pi\sum_a\phi_a^\dagger T_a^A\phi_a
$, together with the real-mass equations
$\mathsf M_a\phi_a=0$ and the F-term equations of the matched
superpotential; solutions are identified modulo gauge
transformations.
Charged condensates therefore lie on their matter walls and
can cancel a nonzero D-linear coefficient, producing Higgs
or mixed sectors. On the Coulomb slice (i.e. without charged condensates), integrating out
retained charged fields instead adds their threshold
contributions back to $F_i^{\mathrm{in}}$, recovering
$F_i$ and hence the condition $F_i=0$ whenever this integration
is perturbatively controlled.
A sufficient regime of control has massive gauge thresholds
parametrically above $e_0^2$, a regular weakly coupled
low-energy chiral description, and decoupled or independently
controlled residual topological sectors.
Weak coupling at the matching scale alone does not establish
these conditions.
Section~\ref{subsubsec:controlledCH} develops the corresponding
criteria for Coulomb, Higgs, and mixed regions.

Unlike the FI and CS matching coefficients, the kinetic terms
can receive further quantum corrections
\cite{Intriligator:2013lca}.
Where a local abelian vector-only action has a regular positive
kinetic matrix, these corrections change the D-term potential
without changing its supersymmetric zero condition.
The outer-patch estimates below establish this regularity.
At massless walls, the light charged fields must remain explicit; depending on the gauge content, the infrared vacuum structure may be perturbatively inaccessible due to the gauge coupling being relevant in three dimensions.

To compare with partition function asymptotics, define
\begin{equation}
 Q(\boldsymbol{x})=\frac16\sum_\rho
 \overline B_3(\rho\cdot\boldsymbol{x}+q^\chi\cdot\boldsymbol{\xi}).
 \label{eq:Qmain}
\end{equation}
Then the dimensionless outer-patch coefficient is
\begin{equation}
 \cF_j(\boldsymbol{x}):=R_1F_j(\boldsymbol{x}/R_1)
 =\tilde\zeta^{(4)}_j-\partial_{x_j}Q(\boldsymbol{x}).
 \label{eq:QvsF}
\end{equation}
For vanishing FI input, $Q$ is the leading phase function in the
Cardy limit of the supersymmetric index. More generally, with
an admissible FI insertion, the local leading phase is
\begin{equation}
 \exp\!\left[2\pi i\frac{R_3^2}{R_1^2}
 (Q-\tilde{\boldsymbol{\zeta}}^{(4)}\cdot\boldsymbol{x})\right].
 \label{eq:QfromI}
\end{equation}

The three-dimensional counterpart is the leading phase
\begin{equation}
 \exp\!\left[2\pi iR_3^2\left(Q^{(3)}(\boldsymbol{\sigma})
 -\boldsymbol\zeta\cdot\boldsymbol\sigma
 -\tfrac12 k_{ij}\sigma_i\sigma_j\right)\right]
 \label{eq:QtildeFromZ}
\end{equation}
in the large-radius $S^3$ partition function, with
\begin{equation}
 F_j(\boldsymbol\sigma)=\zeta_j+k_{ij}\sigma_i
 -\partial_{\sigma_j}Q^{(3)}(\boldsymbol\sigma).
 \label{eq:QtildeVsF}
\end{equation}

The 4-supercharge relations \eqref{eq:QfromI}, \eqref{eq:QtildeFromZ} are analogous to how in the 8-supercharge context the $\mathbb{R}^4\times S^1$ prepotential arises in the Cardy limit of the $S^4\times S^1$ index \cite{Hosseini:2018uzp,Crichigno:2018adf,Choi:2019miv,Hosseini:2021mnn} (cf. the $\Omega$-deformation analog \cite{Nekrasov:1996cz}) or the $\mathbb{R}^4$ prepotential arises in the large-radius limit of the $S^4$ partition function \cite{Russo:2014nka} (cf. the $\Omega$-deformation analog \cite{Nekrasov:2002qd}).

Appendix~\ref{app:QfromW} shows that the $Q$ function arises also in the leading contribution to the $\mathbb{R}^2\times S^1_{r_1}\times S^1_{r_2}$ twisted superpotential $W$ as $r_1\to\infty$:
\begin{equation}
    W\xrightarrow{r_1\to\infty} -i\frac{r_1}{r_2}(Q-\tilde{\boldsymbol{\zeta}}^{(4)}\cdot\boldsymbol{x}).\label{eq:QfromW_intro}
\end{equation}

\subsection{Terminology and assumptions}\label{subsec:terms}

\paragraph{Gauge invariance and boundary conditions.}
We assume a consistent anomaly-free four-dimensional gauge theory
and twists by allowed quantum symmetries, compatible with the
superpotential and four preserved supercharges. The cancellation
$\sum_\rho\rho_i\rho_j\rho_k=0$ makes $\cF_j$ affine within each
chamber. FI inputs are allowed only for abelian factors of the
four-dimensional gauge group. Weight sums include multiplet labels
and multiplicities. The global form of the gauge group fixes the
holonomy and magnetic lattices.

\paragraph{Weak coupling and the Coulomb slice.}
We require weak coupling at $1/R_1$. In an asymptotically free
Lagrangian this follows from $R_1\Lambda_d\ll1$; for an IR-free
theory the compactification scale must lie below its ultraviolet
cutoff in a weakly coupled regime. The calculation is performed
in this Lagrangian regime, even when the theory flows to an
interacting four-dimensional fixed point. On the Coulomb slice,
charged scalar expectation values vanish and neutral scalars are
fixed to satisfy tree-level F-terms. We assume no additional
quadratic superpotential masses for the charged fields on this
slice. Such masses require a corresponding straightforward modification of the
threshold calculations.

\paragraph{Holonomy space, walls, and branches.}
Absent 4d FI parameters, the space of $\boldsymbol{x}$ may be called the
classical Coulomb branch \cite{Strassler:2003qg} or, in another common terminology, the middle-dimensional holonomy slice of the
classical Coulomb branch which also includes the 3d dual photons. A massless threshold on the classical Coulomb branch occurs when a charged
KK mode has zero real mass at $D=0$; codimension-one thresholds are
called walls. Their complement consists of chambers. At the perturbative quantum level, a positive-dimensional zero locus of the perturbative holonomy potential (including 4d FI parameters) acquires dual-photon partners
only in unlifted abelian directions; together, these partners parametrize the perturbative Coulomb
branch. Isolated zeros of the potential with nonsingular
CS matrix instead describe topological gauge sectors.

\paragraph{Matching and scope.}
Following Ref.~\cite{Ardehali:2021irq}, we introduce a cutoff
$\Lambda_M=\epsilon/R_1$ and retain every mode below it. The
continued outer coefficient is continuous at walls, but its slope
can jump.\footnote{By this continuation we never mean actually extending the outer effective action through a massless
threshold. As explained in Section~\ref{sec:scalar_pot}, the continuation is a useful mathematical device for: $i$) obtaining the inner couplings through averaging, $ii$) diagnosing inner-patch perturbative Coulomb branches.} Subtracting each retained mode gives smooth inner
couplings. At a single threshold the inner CS level is the average
of the neighboring outer levels. At intersections one must subtract
all retained contributions explicitly. Kinetic terms affect the perturbative scalar
potential but not the outer supersymmetric zero condition, as demonstrated in Section~\ref{subsec:outer}. Inner vacua additionally
depend on retained matter, its moment map, and the matched
superpotential. The additional infrared information in the
$(A_1,A_4)$ example is the exception to our perturbative focus.
Nonperturbative F-terms are excluded throughout.

\subsection{Plan of the paper}

Section~\ref{sec:scalar_pot} develops the patchwise Wilsonian
description, derives the D-linear couplings from the
Kaluza--Klein towers, and explains light-field matching and
the limitations of the outer-patch vacuum equations.
Section~\ref{sec:ML} discusses the geometry and numerical
reconstruction of higher-rank zero loci.
Section~\ref{sec:examples} applies these methods to flavor-twisted
$\mathrm{SU}(2)$ $\mathcal{N}=4$ super-Yang--Mills theory and
to higher-rank Argyres--Douglas reductions.
Section~\ref{sec:discussion} summarizes the results and the
directions for further work.
Appendix~\ref{app:index} relates the patchwise D-term equations
to asymptotics of supersymmetric partition functions, including
their three-dimensional counterpart.
Appendix~\ref{app:QfromW} derives the associated function $Q$
from a limit of the twisted superpotential on
$\mathbb{R}^2\times T^2$.

\section{Perturbative D-term potential}\label{sec:scalar_pot}

Consider an anomaly-free 4d $\mathcal{N}=1$ gauge theory on $\mathbb{R}^3\times S^1$. The radius of $S^1$ is $R_1,$ and the rank of the (reductive compact Lie) gauge group $G$ is $r_G$.

We perform Fourier expansion of all  the fields around $S^1$, arriving at a 3d $\mathcal{N}=2$ gauge theory with infinitely many fields. For this Kaluza-Klein approach to be reliable, we need the theory to be weakly coupled at the scale $\Lambda_\text{KK}=1/R_1,$ which we assume. For example, if the 4d theory is asymptotically free with dynamical scale $\Lambda_d\,,$ we take
\begin{equation}
    R_1\ll\frac{1}{\Lambda_d}\,.\label{eq:weak-circle_AF}
\end{equation}
We then adopt as a Wilsonian cutoff
\begin{equation}
    \Lambda_M=\frac{\epsilon}{R_1},
\end{equation}
with $0<\epsilon<1/2$ fixed while the coupling is taken small.
At fixed twists and FI inputs, we choose $\epsilon$ sufficiently small
that every isolated off-wall zero of the perturbative holonomy potential lies in the outer patch. The
sufficient hierarchy $g_4^2(1/R_1)\ll\epsilon^2\ll1$ controls
both the outer-patch kinetic and auxiliary-field expansions as discussed below
Eq.~\eqref{eq:generalDmetric}.

All the 3d fields (obtained from KK expansion) that have absolute real mass greater than $\Lambda_M$ will be integrated out.

An important subset of the 3d fields are the $r_G$ compact scalars arising from the gauge holonomy around $S^1.$ These have zero real-mass and sit inside the same 3d $\mathcal{N}=2$ multiplet as the 3d massless gauge fields of the Cartan $U(1)$s. In an integral cocharacter basis, denote their dimensionless coordinates by $(x_1,\dots,x_{r_G})=:\boldsymbol{x}$ and take $x_j\in(-1/2,1/2]$. These parametrize a torus $T^{r_G}$, which (absent 4d FI parameters and modulo Weyl redundancy) is the 3d $\mathcal{N}=2$ classical Coulomb branch.\footnote{In another common terminology, this is ``a middle-dimensional section'' of the classical Coulomb branch, the other half of which is (modulo singularity-related subtleties) parametrized by dual photons---see Section~\ref{subsec:terms}.}

Take a 4d $\mathcal{N}=1$ chiral multiplet $\chi$ in representation $\mathcal{R}_\chi$ of $G,$ and denote a weight of $\mathcal{R}_\chi$ by $\rho$. Fourier expansion of $\chi$ yields 3d $\mathcal{N}=2$ chiral multiplets with real masses
\begin{equation}
    \frac{1}{R_1}(\rho\cdot {\boldsymbol{x}}+q^\chi\!\cdot \boldsymbol{\xi}+n)\,,\qquad n\in\mathbb{Z}\,.
\end{equation}
Here $q^\chi$ is the flavor group representation weight of $\chi\,,$ and $\boldsymbol{\xi}$ parametrizes the twisted boundary condition of the $\chi$ fields around $S^1$, whose coordinate is denoted $y$: 
\begin{equation}
    \varphi(y+2\pi R_1)=e^{2\pi iq^\chi\cdot \boldsymbol{\xi}}\varphi(y)\,.\label{eq:fTwist}
\end{equation}

Similarly, Fourier expansion of a 4d $\mathcal{N}=1$ vector multiplet associated with roots $\alpha_\pm$ of $G$ yields 3d $\mathcal{N}=2$ vector multiplets with real masses
\begin{equation}
    \frac{1}{R_1}(\alpha_\pm\!\cdot {\boldsymbol{x}}+n)\,,\qquad n\in\mathbb{Z}\,.
\end{equation}

In the Wilsonian approach, we distinguish between
    \begin{equation}
    \begin{split}
        \text{the \textbf{\emph{outer patch}} where}\quad &|\mathrm{dist}(\rho\cdot {\boldsymbol{x}}+q^\chi\!\cdot \boldsymbol{\xi},\,\mathbb{Z})|\ge\epsilon\qquad\text{for all $\rho\neq0$},\\
        \text{and}&\qquad|\mathrm{dist}(\alpha_+\!\cdot {\boldsymbol{x}},\,\mathbb{Z})|\ge\epsilon\qquad\text{for all $\alpha_+$},
        \end{split}
    \end{equation}
which has the minimal set of light fields (none of them charged),
and the rest of the holonomy space, which in turn is decomposed into
\begin{equation}
    \begin{split}
\text{\textbf{\emph{inner patches}} where}\qquad &|\mathrm{dist}(\rho\cdot {\boldsymbol{x}}+q^\chi\cdot \boldsymbol{\xi},\,\mathbb{Z})|<\epsilon\quad\text{for some $\rho\neq0$},\\
        \text{or}&\qquad|\mathrm{dist}(\alpha_+\!\cdot {\boldsymbol{x}},\,\mathbb{Z})|<\epsilon\quad\text{for some $\alpha_+$}.
    \end{split}
\end{equation}
The symbol dist($x,\,\mathbb{Z}$) stands for the signed distance of $x$ from $\mathbb{Z}$, namely $x-\mathrm{nint}(x)$, with nint the nearest integer function.

In short, inner patches are where there are light \emph{charged} multiplets in the 3d KK EFT, with light meaning absolute real mass smaller than the Wilsonian cutoff $\Lambda_M=\frac{\epsilon}{R_1}.$
Inner patches are neighborhoods of the matter and vector threshold loci, whose points $\boldsymbol{x}^p$ and $\boldsymbol{x}^g$ satisfy
\begin{equation}
\begin{split}
         \rho\cdot {\boldsymbol{x}^p}+q^\chi\cdot \boldsymbol{\xi}\in\mathbb{Z}\quad&\text{for some $\rho_\chi\neq0\,$},\\
         \alpha_+\cdot {\boldsymbol{x}^g}\in\mathbb{Z}\quad&\text{for some $\alpha^{}_+$\,},
         \end{split}\label{eq:walls}
    \end{equation}
where additional massless fields associated to $\rho$ and $\alpha_\pm$ arise. We denote the locations $\{\boldsymbol{x}^{p_1},\dots,\boldsymbol{x}^{g_1},\dots\}$ collectively by $\{\boldsymbol{x}^{s_1},\boldsymbol{x}^{s_2},\dots\}.$ The codimension-1 hyperplanes \eqref{eq:walls} are the walls inside the classical Coulomb branch. See Figure~\ref{fig:U1-SU2}, where on the left there is a wall at $x=0,$ and on the right a wall at $x=0$ and another at $x=\pm\frac{1}{2}$.

\subsection{The outer patch}\label{subsec:outer}

There are no charged fields in the 3d EFT governing the outer-patch. The light degrees of freedom comprise a 3d $\mathcal{N}=2$ $U(1)^{r_G}$ gauge theory, possibly with uncharged chiral multiplets. With a tree kinetic matrix $\mathcal G^{(0)}_{ij}=e_0^{-2}\delta_{ij}$, the leading perturbative D-term potential on the Coulomb slice is
\begin{equation}
    V_D^{(0)}=\frac{e_0^2}{32\pi^2} \sum_j F_j({\boldsymbol{\sigma}})  F_j({\boldsymbol{\sigma}})\,,\label{eq:Vsc_F}
\end{equation}
where, as we reviewed in Section~\ref{sec:intro}:
\begin{equation}
F_j(\boldsymbol{\sigma}):={\zeta}^\text{eff}_j+k_{ij}^\text{eff}\sigma_i\,.\label{eq:Fj-def}
\end{equation}

We now use (cf.~Eq.~(2.6) of \cite{Intriligator:2013lca})\footnote{Dimensional reduction relates the dimensionful FI parameters as $\zeta:=\zeta^{(3)}=4\pi^2 R_1\zeta^{(4)}.$ With the dimensionless 4d FI parameter defined through $\tilde{\zeta}^{(4)}=(2\pi R_1)^2\,\zeta^{(4)},$ we get the relation $\zeta=\frac{1}{R_1}\tilde{\zeta}^{(4)}$.}
\begin{equation}
\begin{split}
{\zeta}^\text{eff}_j={\zeta}_j+{\zeta}^\text{1-loop}_j&=\frac{1}{R_1}\tilde{\zeta}^{(4)}_j+\sum_{n\in\mathbb{Z}}\frac{1}{2}\sum_{\rho}\rho_j\,\frac{n+q^\chi\cdot\boldsymbol{\xi}}{R_1}\,\mathrm{sign}(n+\rho\cdot\boldsymbol{x}+q^\chi\cdot\boldsymbol{\xi})\,,\\
k_{ij}^\text{eff}\sigma_i=k_{ij}^\text{1-loop}\sigma_i&=\sum_{n\in\mathbb{Z}}\frac{1}{2}\sum_{\rho}\rho_j\,\frac{\rho\cdot\boldsymbol{x}}{R_1}\,\mathrm{sign}(n+\rho\cdot\boldsymbol{x}+q^\chi\cdot\boldsymbol{\xi})\,.
\end{split}\label{eq:zeta_k_eff}
\end{equation}
Paired vector modes give canceling contributions to these, although they correct the
kinetic matrix and determine vector thresholds.

A case of particular interest is when 
\begin{equation}
    \boldsymbol{\xi}=\boldsymbol{\xi}^0+R_1\boldsymbol{m}^{}_f\,,\label{eq:AZ}
\end{equation}
with $\boldsymbol{\xi}^0=\mathcal{O}(R_1^0)$ \cite{Amariti:2024bdd,Amariti:2025gca}, where $\boldsymbol{m_f}$ can be thought of as a finite real-mass in the 3d KK EFT. In this case, one may choose to keep the definition of ${\zeta}^\text{1-loop}_j$ as above, or may want to separate from it the piece proportional to $\boldsymbol{m}_f$:
\begin{equation}
    {\zeta}^\text{1-loop}_j\to {\zeta}^\text{1-loop}_j+k^\text{1-loop}_{j\ell}{{m_f}}^\ell,\label{eq:xi_plus_kjfmf}
\end{equation}
where $\ell$ labels the $U(1)$ Cartans of the flavor group. We choose the former option and denote the whole combination on the RHS of \eqref{eq:xi_plus_kjfmf} by ${\zeta}^\text{1-loop}_j$.

We now compute
\begin{equation}
\begin{split}
    F_j(\boldsymbol{\sigma})&=\frac{1}{R_1}\tilde{\zeta}^{(4)}_j+\frac{1}{2R_1}\sum_{\rho}\rho_j\big(\!\sum_{n\in\mathbb{Z}}\mathrm{sign}(n\!+\!\rho\cdot\boldsymbol{x}\!+\!q^\chi\cdot\boldsymbol{\xi})\,(n\!+\!\rho\cdot\boldsymbol{x}\!+\!q^\chi\cdot\boldsymbol{\xi})\big)\\
    &=\frac{1}{R_1}\Big(\tilde{\zeta}^{(4)}_j-\frac{1}{2}\sum_{\rho}\rho_j\overline{B}_2(\rho\cdot\boldsymbol{x}\!+\!q^\chi\cdot\boldsymbol{\xi})\Big)\,,
    \end{split}\label{eq:Fj}
\end{equation}
where the second line uses the regulated identity
\begin{equation}
 \sum_{n\in\mathbb Z}^{\zeta}|n+u|
 =\zeta(-1,\{u\})+\zeta(-1,1-\{u\})=-\overline B_2(u),
 \label{eq:KKsum}
\end{equation}
with values at integers defined by continuity. Here
$\overline B_j(u)=B_j(\{u\})$
and $B_j$ are the Bernoulli polynomials. In particular $\overline{B}_2(x):=B_2(\{x\})=-\{x\}(1-\{x\})+\frac{1}{6}.$

The globally matched leading potential on the outer patch is therefore
\begin{equation}
    \boxed{V_D^{(0)}=\frac{e_0^2}{32\pi^2}|\vec F(\boldsymbol{\sigma})|^2=\frac{e_0^2}{32\pi^2R_1^2}\,\Big|\vec{\tilde{\zeta}}^{(4)}-\frac{1}{2}\sum_{\rho}\vec\rho\ \overline{B}_2(\rho\cdot\boldsymbol{x}\!+\!q^\chi\cdot\boldsymbol{\xi})\Big|^2.}\label{eq:Vsc}
\end{equation}

\subsection*{Exact relations for outer vacua}

Away from matter walls,
\begin{equation}
 k^{\mathrm{out}}_{ij}=\partial_{x_i}\cF_j
 =-\sum_\rho\rho_i\rho_j\overline B_1
 (\rho\cdot\boldsymbol{x}+q^\chi\cdot\boldsymbol\xi).
 \label{eq:kout}
\end{equation}
Here $\overline B_1(u)=\{u\}-1/2$ off integers. Gauge-anomaly cancellation makes this matrix
constant on each chamber. The FI couplings can be found from \eqref{eq:Fj-def}, \eqref{eq:Fj}, and \eqref{eq:kout}. The FI and CS matching coefficients
are one-loop exact in the fixed scheme, unlike the kinetic matrix
\cite{Intriligator:2013lca}. Consequently the outer D-flat condition
would follow from \eqref{eq:Vsc} to be
\begin{equation}
\tilde{\zeta}^{(4)}_j-\frac{1}{2}\sum_{\rho}\rho_j\ \overline{B}_2(\rho\cdot\boldsymbol{x}\!+\!q^\chi\cdot\boldsymbol{\xi})=0\,,\quad\text{for $j=1,\dots,r_G,$}
\label{eq:vacEqs}
\end{equation}
if we could establish a regular positive kinetic matrix. This is done through uniform control estimates on the outer patch next.

Note that the outer patch has in general multiple connected components separated by the walls \eqref{eq:walls}. Each connected component of the outer patch is described by a 3d EFT with its own effective couplings. The equations \eqref{eq:Vsc} and \eqref{eq:vacEqs} apply on all connected components uniformly, and yield the perturbative vacua of each 3d EFT.

\subsection*{Establishing uniform control}

Equations~\eqref{eq:Vsc_F} and \eqref{eq:Vsc} use the leading
Maxwell kinetic term. More generally,
\begin{equation}
 \mathcal L_D=\tfrac12D_i\mathcal G_{ij}D_j
 -\frac{F_iD_i}{4\pi}+O(D^3),\qquad
 V_{D,\mathrm{quad}}=\frac{F_i(\mathcal G^{-1})_{ij}F_j}{32\pi^2}.
 \label{eq:generalDmetric}
\end{equation}
A general positive tree kinetic matrix replaces
$e_0^2\delta_{ij}$ by $(\mathcal G^{(0)})^{-1}_{ij}$ in the leading
potential. At zero FI input, $F_i$ is generated at one loop and
its square gives a two-loop-order energy of order $g_4^2/R_1^3$.
Tree-level 4d FI inputs make additional contributions.

There are two distinct control conditions needed for the reliability of \eqref{eq:Vsc}: small corrections to
the quadratic kinetic terms, and validity of the expansion in
the auxiliary fields. Both can be understood from a massive
charged chiral multiplet. For a mode of charges $Q_i$ and real
mass $M$, the off-shell scalar Lagrangian contains
\begin{equation}
 \mathcal L_\phi\supset
 -M^2|\phi|^2
 +\frac12(\boldsymbol Q\cdot\boldsymbol D)|\phi|^2,
 \qquad
 m_\phi^2(\boldsymbol D)
 =M^2-\frac12\boldsymbol Q\cdot\boldsymbol D.
\end{equation}
The second term follows by expanding the gauge-covariant
matter kinetic term in superspace; see
\cite[Eqs.~(A.11)--(A.14)]{Intriligator:2013lca}.
Thus a background auxiliary field shifts the mass
squared. (The factor $1/2$ is fixed by the auxiliary-field
normalization used here and does not affect the following
parametric estimates.)

The one-loop correction to the coefficient of $D_iD_j$ is
obtained from a scalar loop with two insertions of this
interaction:
\begin{center}
\begin{tikzpicture}[
  line width=0.8pt,
  scalar/.style={
    postaction={decorate},
    decoration={
      markings,
      mark=at position 0.5 with {\arrow{>}}
    }
  }
]
  \draw[dashed] (-2.2,0) node[left] {$D_i$} -- (-0.8,0);
  \draw[dashed] ( 0.8,0) -- (2.2,0) node[right] {$D_j$};

  \draw[scalar] (-0.8,0) arc[start angle=180,end angle=0,radius=0.8];
  \draw[scalar] ( 0.8,0) arc[start angle=0,end angle=-180,radius=0.8];

  \fill (-0.8,0) circle[radius=1.6pt];
  \fill ( 0.8,0) circle[radius=1.6pt];

  \node at (0, 1.08) {$\phi$};
  \node at (0,-1.08) {$\phi$};
\end{tikzpicture}
\end{center}
The vertices supply the charge factor $Q_iQ_j$.
The zero-external-momentum integral is
\begin{equation}
 I_2(M):=
 \int\frac{d^3p}{(2\pi)^3}\frac{1}{(p^2+M^2)^2}
 =\frac{1}{8\pi|M|}.
\end{equation}
Consequently a massive chiral mode contributes
$\delta\mathcal G_{ij}=O(Q_iQ_j/|M|)$.
Supersymmetry places this correction in the same local
two-derivative action as the vector-multiplet kinetic terms.\footnote{Equivalently, the kinetic matrix is proportional to the
Hessian of the function $f$ appearing in the quantum corrected gauge kinetic term as
$-\int d^4\theta\,f(\boldsymbol\Sigma)$ \cite[Eqs.~(A.15)--(A.16),
Eqs.~(B.1)--(B.2)]{Intriligator:2013lca}; differentiating
the one-loop logarithmic term in $f$ produces the same
inverse-mass dependence.
The general charge-dependent expression is displayed in
\cite[Eq.~(2.20)]{deBoer:1997kr}.}
Massive vector multiplets give the same inverse-mass scaling
with different coefficients and signs; see
\cite[Eqs.~(2.21)--(2.22)]{deBoer:1997kr}.
These are one-loop threshold estimates;
the kinetic matrix of course receives higher-loop corrections as well.

For the KK tower, the ultraviolet part of the kinetic
correction is absorbed into the renormalized gauge coupling.
The remaining threshold dependence is bounded on the outer
patch using $|M|\geq\Lambda_M=\epsilon/R_1$.
With field content and charges fixed, this gives
$\|\delta\mathcal G\|=O(R_1/\epsilon)$ and hence, relative
to $\mathcal G^{(0)}=e_0^{-2}\mathbf 1$,
\begin{equation}
 \bigl\|(\mathcal G^{(0)})^{-1}\delta\mathcal G\bigr\|
 =O\!\left(\frac{e_0^2R_1}{\epsilon}\right)
 =O\!\left(\frac{g_4^2}{\epsilon}\right).
\end{equation}
Smallness of this correction controls the kinetic matrix,
but a separate condition is needed to truncate the heavy
determinants at quadratic order in $D$.
Indeed, successive auxiliary-field insertions in the scalar
propagator introduce powers of
$(\boldsymbol Q\cdot\boldsymbol D)/(2(p^2+M^2))$.
A sufficient condition for their uniform suppression is
therefore
$|\boldsymbol Q\cdot\boldsymbol D|\ll M^2$.
It also ensures that the auxiliary background does not drive
a heavy scalar close to a massless or tachyonic threshold.

At the quadratic saddle
$D_i=(\mathcal G^{-1})_{ij}F_j/(4\pi)$, the small metric
correction implies $\|\mathcal G^{-1}\|\sim\|(\mathcal G^{(0)})^{-1}\|=O(e_0^2)$.
Using $F_j=\cF_j/R_1$ and $M^2\geq\epsilon^2/R_1^2$,
the auxiliary-field expansion parameter is thus bounded,
up to fixed charge and numerical factors, by
\begin{equation}
 \frac{|\boldsymbol D|}{M^2}\sim\frac{g_4^2}{\epsilon^2}
 \sup_{\mathrm{out}}|\boldsymbol{\cF}(\boldsymbol{x})|.
 \label{eq:Dcontrol}
\end{equation}
At fixed twists, dimensionless FI inputs, and field content,
$\boldsymbol{\cF}$ is bounded. The sufficient hierarchy
$g_4^2\ll\epsilon^2\ll1$ therefore controls both expansions
uniformly and keeps $\mathcal G$ positive and nonsingular.
At a supersymmetric zero, $D=0$, the auxiliary-field
condition is automatic. Note that the cutoff ($\propto\epsilon$) cannot be sent to zero
at fixed coupling while maintaining these estimates.

\subsection{Inner patches: matching and vacuum conditions}\label{subsec:inner}

An inner patch retains the charged modes with $|M|<\Lambda_M$.
We compute its Wilsonian couplings and state the vacuum equations
and control criteria for Coulomb, Higgs, and mixed regions.

\subsubsection{Subtracting retained modes}\label{subsubsec:retained}

For $\epsilon<1/2$, each chiral weight has at most one retained
KK mode. Write $u_\rho=\rho\cdot\boldsymbol{x}
+q^\chi\cdot\boldsymbol\xi$ and
$\delta(u):=\mathrm{dist}(u,\,\mathbb{Z})=u-\mathrm{nint}(u)$. Let $L_{\mathrm{in}}$ label
the retained chiral modes, including multiplicities, and
$H_{\mathrm{in}}$ its complement in the weight sum. The retained
mass is $M_\rho=\delta(u_\rho)/R_1$. Replacing
\begin{equation}
 \overline K_2(u)=\overline B_2(u)+|\delta(u)|,\qquad
 \overline K_1(u)=\overline B_1(u)+\tfrac12\mathrm{sign}\delta(u)
 \label{eq:Kfunctions}
\end{equation}
removes precisely this mode from its KK tower
\cite{Ardehali:2021irq}. We set $\mathrm{sign}(0)=0$.
The dimensionful inner coefficient is
\begin{equation}
\boxed{F^{\mathrm{in}}_j(\boldsymbol\sigma)=\frac1{R_1}\bigg[
 \tilde\zeta^{(4)}_j
 -\frac12\sum_{\rho\in H_{\mathrm{in}}}\rho_j\overline B_2(u_\rho)
 -\frac12\sum_{\rho\in L_{\mathrm{in}}}\rho_j\overline K_2(u_\rho)
 \bigg].}
\label{eq:Fj-in}
\end{equation}
Near an integer, $|\delta|<1/2$ gives
\begin{equation}
 \overline B_2(u)=\delta^2-|\delta|+\tfrac16,\qquad
 \overline K_2(u)=\delta^2+\tfrac16.
 \label{eq:BvsK}
\end{equation}
Thus the inner coefficient is smooth across every retained
threshold. Gauge-anomaly cancellation again makes it affine.
At a reference wall point $\boldsymbol{x}_*$, set
$\boldsymbol{s}=(\boldsymbol{x}-\boldsymbol{x}_*)/R_1$ and write
\begin{equation}
\begin{split}
 F^{\mathrm{in}}_j(\boldsymbol{x}_*/R_1+\boldsymbol{s})
 &=\zeta_{j,\mathrm{loc}}+k^{\mathrm{in}}_{ij}s_i,\\
 \zeta_{j,\mathrm{loc}}&=F^{\mathrm{in}}_j(\boldsymbol{x}_*/R_1),\\
 k^{\mathrm{in}}_{ij}
 &=-\sum_{\rho\in H_{\mathrm{in}}}\rho_i\rho_j\overline B_1(u_\rho)
 -\sum_{\rho\in L_{\mathrm{in}}}\rho_i\rho_j\overline K_1(u_\rho).
\end{split}
\label{eq:innercouplings}
\end{equation}
These are supersymmetric Wilsonian CS levels. At vector walls,
the retained vector multiplets must also be restored and the
Cartan data assembled into couplings of the unbroken gauge group,
with the global form inherited from four dimensions.

The exact coefficient matching is
\begin{equation}
\begin{split}
 F_j(\boldsymbol\sigma)&=F^{\mathrm{in}}_j(\boldsymbol\sigma)
 +\tfrac12\sum_{\rho\in L_{\mathrm{in}}}\rho_j|M_\rho|,\\
 k^{\mathrm{out}}_{ij}&=k^{\mathrm{in}}_{ij}
 +\tfrac12\sum_{\rho\in L_{\mathrm{in}}}
 \rho_i\rho_j\mathrm{sign}(M_\rho).
\end{split}
\label{eq:thresholdmatching}
\end{equation}
The second identity holds off the walls. If every retained chiral
is massless at $\boldsymbol{x}_*$, the two D-coefficients agree
there. They need not have the same zero set on the surrounding
patch. At intersecting walls, each
retained contribution is subtracted independently. Varying the
cutoff redistributes contributions between explicit fields and
couplings without changing the recombined data.

\subsubsection{Controlled Coulomb and Higgs sectors}
\label{subsubsec:controlledCH}

A sufficient criterion for local control of the matched inner
theory is that its interacting gauge sector can be integrated
out perturbatively, leaving a weakly coupled theory of chiral
multiplets. On Coulomb or mixed branches, these may include
chirals dual to surviving unlifted abelian vectors, once all
fields charged under them are massive
\cite[Appendix B]{Intriligator:2013lca}.
The resulting K\"ahler description must be regular and
perturbatively controlled, with weak superpotential interactions
at the energies under consideration.
Any residual topological gauge sector must decouple from
these chirals or be independently controlled.\footnote{At
fixed level, a Chern--Simons mass is of order $e_0^2$ and
does not by itself establish perturbative decoupling from
light charged matter.}
Establishing this criterion requires checking both the
mass hierarchy and the retained-field corrections to the
vacuum equations.

\paragraph{Matched vacuum equations.}
Let $H$ be the full wall gauge group, choose an orthonormal
basis of Hermitian generators $T^A$, and denote their action
on the retained matter representations $R_a$ by $T_a^A$.
Write $\mathsf s=s^A T^A$ for the vector-multiplet scalar
measured from the reference wall.
Suppressing the reference-point shift in its argument,
the covariant form of Eq.~\eqref{eq:innercouplings} is
\[
 F_A^{\mathrm{in}}(\mathsf s)
 =\zeta_{A,\mathrm{loc}}+k_{AB}^{\mathrm{in}}s^B,
\]
where $\zeta_{\mathrm{loc}}$ is supported on the abelian
center and $k^{\mathrm{in}}$ is an invariant symmetric
bilinear.
With canonical matter kinetic terms, the classical
supersymmetric equations of the matched action are
\cite{Intriligator:2013lca}
\begin{equation}
 \begin{aligned}
 2\pi\sum_a\phi_a^\dagger T_a^A\phi_a
 &=F_A^{\mathrm{in}}(\mathsf s),\\
 \mathsf M_a(\mathsf s)\phi_a&=0,
 \qquad \partial_{\Phi_I}W_{\mathrm{loc}}=0.
 \end{aligned}
 \label{eq:wallvacua}
\end{equation}
Here $\mathsf M_a(\mathsf s)=\mathsf M_a(0)+s^A T_a^A$,
with $\mathsf M_a(0)$ commuting with $H$, and $\Phi_I$ runs
over all retained chiral components, including neutral ones.
Solutions are quotiented by $H$; for noncanonical matter
kinetics, the corresponding moment map replaces the
quadratic sum. The abelian equations follow by setting
$T_a^i=\rho_{a,i}$.
The matched $W_{\mathrm{loc}}$ includes tree-level elimination
of heavy chirals but no perturbatively generated superpotential
terms \cite{Grisaru:1979wc,Seiberg:1993vc,Weinberg:1998uv};
nonperturbative F-terms remain outside our scope.
These equations are classical only in the retained fields:
their couplings already include the modes removed at matching,
while retained-field loops must still be included or shown
negligible.

For condensates $v_a=\langle\phi_a\rangle$ that are massless
at the reference wall, $\mathsf M_a(0)v_a=0$, the real-mass
equation requires $s^A T_a^A v_a=0$.
In a Cartan gauge this becomes
\[
 (\boldsymbol\rho\cdot\boldsymbol{s})v_{a,\rho}=0.
\]
Thus every condensed nonzero weight lies on its matter wall.
Such a weight fixes $s=0$ in rank one; in higher rank, a
common charge kernel can leave mixed Higgs--Coulomb directions,
subject to the remaining vacuum equations and Chern--Simons
couplings.

\paragraph{Gauge thresholds.}
The gauge masses must be determined using the full Lie algebra
of $H$, including its off-Cartan generators.
With the leading equal-coupling normalization,
\[
 \mathcal L_{\mathrm{kin}}\supset
 -\frac{1}{4e_0^2}F_{\mu\nu}^A F^{A\mu\nu}
 -\frac{1}{2e_0^2}\|D_\mu\mathsf s\|^2
 -\sum_a|D_\mu\phi_a|^2,
\]
canonical normalization $A_\mu=e_0 A_{\mu,\mathrm c}$ gives,
on constant scalar backgrounds,
\[
 D_\mu v_a=-ie_0 A_{\mu,\mathrm c}^A T_a^A v_a,
 \qquad
 D_\mu\mathsf s=e_0 R_{\mathsf s}(A_{\mu,\mathrm c}),
 \qquad
 R_{\mathsf s}(X)\equiv-i[X,\mathsf s].
\]
Substitution yields
\begin{equation}
 \begin{aligned}
 \mathcal L_{\mathrm{mass}}^{(2)}
 &=-\frac12 A_{\mu,\mathrm c}^A
 (m_V^2)_{AB}A_{\mathrm c}^{B\mu},\\
 (m_H^2)_{AB}
 &=2e_0^2\operatorname{Re}\sum_a
 (T_a^A v_a)^\dagger(T_a^B v_a),\\
 m_V^2&=m_H^2+R_{\mathsf s}^{\mathsf T}R_{\mathsf s},
 \end{aligned}
 \label{eq:innerHiggsmatrix}
\end{equation}
where the transpose uses the chosen invariant inner product.
Both contributions are positive semidefinite, so the kernel
consists precisely of the unbroken generators:
\[
 T_a(X)v_a=0\quad\text{for every }a,
 \qquad [X,\mathsf s]=0.
\]
In a Cartan gauge,
$[\mathsf s,E_\alpha]=(\alpha\cdot\boldsymbol{s})E_\alpha$,
so the adjoint contribution has eigenvalue
$|\alpha\cdot\boldsymbol{s}|^2$ on each root plane.
It vanishes at $\mathsf s=0$, although matter condensates
can still give masses to both Cartan and off-Cartan fields.
In general, the full matrix must be diagonalized to identify
the massive combinations and the unbroken gauge algebra.

Since $e_0^2$ has dimensions of mass in three dimensions,
the gauge-loop expansion parameter at a threshold
$m_{V,\ell}>0$ is of order $e_0^2/m_{V,\ell}$.
At fixed matter content and Chern--Simons levels, a sufficient
hierarchy for perturbative integration of the massive gauge
sector is therefore%
\footnote{Higgs masses need not lie below the Coulomb-branch
matching cutoff $\Lambda_M$. Thresholds above this cutoff
require matching in the condensed background, using a
description valid at those masses and matching its couplings
and kinetic terms accordingly. Omitted fields must remain
heavy in that background; any that become light must be
reinstated.}
\begin{equation}
 e_0^2\ll m_{V,\ell},
 \qquad
 m_{V,\ell}^2\in\operatorname{Spec}_{>0}(m_V^2).
 \label{eq:wallhierarchy}
\end{equation}
The eigenvalues of $m_V^2$ are parity-even mass parameters.
Chern--Simons terms modify the physical pole masses, but at
fixed levels these modifications are subleading under
Eq.~\eqref{eq:wallhierarchy}.
Surviving free abelian photons require no such hierarchy, as mentioned at the beginning of this subsubsection.
Any residual interacting gauge sector requires separate
control.

\paragraph{Retained-field corrections.}
We recall the standard one-loop D-tadpole, which for a chiral
with only a real mass gives precisely the Wilsonian threshold
already used in Eq.~\eqref{eq:thresholdmatching}
\cite[Eq.~(2.6)]{Intriligator:2013lca}.
Here we apply the same calculation to fields retained at the
initial matching scale, distinguishing real-mass thresholds
from corrections around a Higgs condensate. For an abelian gauge direction $i$, expanding
$\phi_a=v_a+\eta_a$ gives a vertex proportional to
$\rho_{a,i}D_i|\eta_a|^2$.
Contracting the fluctuation fields produces a tadpole:
\begin{center}
\begin{tikzpicture}[
  line width=0.8pt,
  scalar/.style={
    postaction={decorate},
    decoration={
      markings,
      mark=at position 0.5 with {\arrow{>}}
    }
  }
]
  \draw[dashed] (-2.2,0) node[left] {$D_i$} -- (-0.8,0);
  \draw[scalar] (-0.8,0)
    arc[start angle=180,end angle=0,radius=0.8];
  \draw[scalar] (0.8,0)
    arc[start angle=0,end angle=-180,radius=0.8];
  \fill (-0.8,0) circle[radius=1.6pt];
  \node at (0,1.08) {$\eta_a$};
\end{tikzpicture}
\end{center}
For a scalar propagator of mass $m\geq0$, the same
mass-independent FI subtraction used in matching gives
\[
 I(m)-I(0)
 =
 \int\frac{d^3p}{(2\pi)^3}
 \left(\frac{1}{p^2+m^2}-\frac{1}{p^2}\right)
 =
 -\frac{m}{4\pi}.
\]
The finite correction is therefore set by the mass in the
fluctuation propagator.

For a noncondensing chiral whose only mass is the real mass
$M_\rho$, setting $m=|M_\rho|$ reproduces the familiar threshold
\[
 \Delta F_i=\frac12\rho_i|M_\rho|.
\]
When the chiral is subsequently integrated out, assuming $e_0^2\ll|M_\rho|$, this
contribution enters the Wilsonian D-coefficient, correcting the right-hand side of \eqref{eq:wallvacua}. This threshold contains no explicit factor of $e_0^2$:
weak coupling makes its calculation reliable but does not
make it small relative to the terms determining the vacuum.
It must therefore be included unless it is independently
negligible.

For a condensed charged component, the situation is different. A condensed charged component satisfies
$M_\rho(\boldsymbol{s})=0$ at the classical level and hence supplies no independent
real-mass threshold at leading order.
Its fluctuations can nevertheless acquire Higgs-scale
masses through gauge interactions, including the quadratic
terms obtained by expanding the D-term potential.
Using these massive propagators already incorporates the
interactions responsible for their masses, even though the
tadpole has no explicit gauge propagator.
The correction thus need not vanish, but its control criteria are straightforward to establish.

For example, at a regular abelian Higgs vacuum with
characteristic condensate magnitude $v$ and
$m_H\sim e_0v$, the gauge-generated correction to the
matter moment map scales as
\[
 \delta\mu_H
 =O\!\left(\frac{m_H}{4\pi}\right)
 =O\!\left(\frac{e_0^2}{4\pi m_H}v^2\right),
 \qquad
 \frac{\delta\mu_H}{v^2}
 =O\!\left(\frac{e_0^2}{4\pi m_H}\right),
\]
with fixed charge and multiplicity factors understood.
Its coefficient requires the coupled fluctuation spectrum.
These corrections are encoded in the matter K\"ahler terms,
which determine both the kinetic terms and the moment map.\footnote{Shizuya's abelian Higgs model makes this connection explicit
\cite[Sec.~V, Eqs.~(5.2)--(5.3), and App.~A]{Shizuya:2005th}.
After subtracting the FI divergence, the finite term in
Eq.~(5.2) gives the one-loop matter K\"ahler potential,
in his normalization,
\[
 K_{\mathrm{eff}}(X)
 =X-\frac{1}{2\pi}\sqrt{2e^2X},
 \qquad X=\bar\Phi e^{2eV}\Phi,
\]
at leading order in the superspace derivative expansion.
Writing $x=|\phi|^2$, the scalar kinetic coefficient and
the matter moment map are respectively
$g_{\phi\bar\phi}
 =K_{\mathrm{eff}}'(x)+xK_{\mathrm{eff}}''(x),\,
 \mu(x)=xK_{\mathrm{eff}}'(x).$ Thus the same K\"ahler correction modifies both quantities. At a condensate $|\phi|=v$, this gives
\[
 \mu(v^2)=v^2-\frac{m}{4\pi},
 \qquad m=\sqrt{2}\,ev,
\]
and the corrected D-equation is $\kappa^{\mathrm{there}}_{\mathrm{ren}}/e_{\mathrm{there}}=\zeta^{\mathrm{here}}_{\mathrm{ren}}/2\pi=v^2-m/4\pi$,
precisely Eq.~(5.3).
The finite correction is field dependent, rather than
a renormalization of the constant FI parameter.
The chiral--vector mixing is displayed in
Eq.~(A5) and the square-root term is derived from the
fluctuation determinant in Eq.~(A8).
This explicitly exhibits a nonzero Higgs-scale correction
despite the vanishing classical real mass, with relative size
$\frac{m}{4\pi v^2}=\frac{e^2}{2\pi m}$, confirming the suppression
estimated above.}
For $e_0^2\ll m_H$, they give only a small displacement of
a regular Higgs solution in its massive directions. For nonabelian backgrounds or several thresholds, the
corresponding gauge-loop check uses the full
matrix~\eqref{eq:innerHiggsmatrix} and
hierarchy~\eqref{eq:wallhierarchy}.

Both estimates require the appropriate fluctuation spectrum.
In particular, the Hessian
$\partial_{\Phi_I}\partial_{\Phi_J}W_{\mathrm{loc}}$
can generate masses even for components with vanishing
real mass, and its loop effects require a separate check
of the superpotential interactions.
Additional masses or mixing require the corresponding
coupled fluctuation determinant rather than the simple
real-mass threshold above (cf.~\cite[Eqs.~(A5)--(A8)]{Shizuya:2005th}).
Fields that remain light must be retained in the low-energy
analysis.
Thus Eq.~\eqref{eq:wallvacua}, with the matched D-coefficient
alone, gives the leading vacuum equations only when these
retained-field corrections are negligible.

\paragraph{Coulomb and mixed directions.}
On the Coulomb slice, charged matter backgrounds vanish,
so retained-field thresholds can determine D-flatness
even at weak coupling.
Fix neutral fields at supersymmetric backgrounds and assume
there are no additional charged quadratic superpotential
masses.
If every retained charged chiral and every restored
off-Cartan vector can be integrated out perturbatively,
\[
 e_0^2\ll|M_\rho|,
 \qquad
 e_0^2\ll|\alpha\cdot\boldsymbol{s}|,
\]
the D-coefficient below these thresholds is
\begin{equation}
 \begin{aligned}
 F_i^{\mathrm C}(\boldsymbol{s})
 &:=F_i^{\mathrm{in}}(\boldsymbol{x}_*/R_1+\boldsymbol{s})
   +\frac12\sum_{\rho\in L_{\mathrm{in}}}
      \rho_i|M_\rho(\boldsymbol{s})|\\
 &=F_i(\boldsymbol{x}_*/R_1+\boldsymbol{s})
  =\frac{\cF_i(\boldsymbol{x}_*+R_1\boldsymbol{s})}{R_1}.
 \end{aligned}
 \label{eq:innerCoulomb}
\end{equation}
The real-mass thresholds can compete with or cancel the
matched coefficient, as discussed above
\cite{Intriligator:2013lca}.
Paired vector contributions cancel in this coefficient
but affect the kinetic terms.
With a regular positive kinetic matrix, D-flatness is
therefore $F_i^{\mathrm C}=F_i=0$, even when
$F_i^{\mathrm{in}}\ne0$.

The same distinction applies to unbroken gauge directions
on a mixed branch.
Since $T_a(X)v_a=0$ for every unbroken generator $X$, the
condensates contribute nothing to those classical
D-equations.
There is therefore no condensate term in these directions
against which a loop correction can be declared small.
Their D-equations must use the appropriate loop-corrected
coefficients and any remaining light-field contributions.

The Letter \cite{CompanionPRL} illustrates cancellation
of the matched D-coefficient both by a condensate and
by retained-field thresholds.
Together with the regular, weakly coupled chiral description
specified above, these mechanisms give local perturbative
control of Higgs, mixed, and Coulomb regions.
Massless interacting charged sectors with light gauge fields
require an independent analysis of the matched theory's
infrared dynamics.

\subsection{Summary: perturbative vacuum sectors}

For fixed external parameters choose $\epsilon$ small enough
that all isolated off-wall zeros lie in the outer patch, then
take $g_4^2\ll\epsilon^2$. These zeros have nonsingular
$k^{\mathrm{out}}$ and define perturbative topological gauge
sectors. A zero with a CS kernel extends locally along that
kernel; together with the corresponding dual photons, these
directions give a perturbative Coulomb branch. Neutral spectator
moduli, if present, must be included separately.

Inner patches can contain controlled extensions of Coulomb
branches, semiclassical Higgs sectors, or mixed branches.
Their vacuum equations involve the matched couplings and
retained fields, with the control criteria of
Section~\ref{subsubsec:controlledCH}. At massless interacting
loci, matching supplies the local theory but does not by itself
establish the existence of supersymmetric vacua, which can be a strong-coupling IR question.

\section{Chamber geometry and numerical reconstruction}\label{sec:ML}

The uniform control established in Section~\ref{subsec:outer} for the outer potential allows capturing the perturbative outer-patch vacua through the system \eqref{eq:vacEqs}.
Alternatively, we can study the zero set of the dimensionless scalar potential
\begin{equation}
 \bar V_D(\boldsymbol{x})=|\boldsymbol{\cF}(\boldsymbol{x})|^2
 =\left|\vec{\tilde\zeta}^{(4)}-\frac12\sum_\rho\vec\rho\,
 \overline B_2(\rho\cdot\boldsymbol{x}+q^\chi\cdot\boldsymbol\xi)\right|^2.
 \label{eq:normalizedVsc}
\end{equation}
We also know that if the zero set of $\bar V_D$ extends from an outer patch to a neighboring inner patch, then it extends through the controlled region of the inner patch as well (cf.~\eqref{eq:innerCoulomb}). In the weak-coupling regime $g_4\to0$ of our interest in this work, the uncontrolled regions of the holonomy space localize to the walls. The natural domains of $\bar V_D$ are therefore the chambers (i.e. the connected components of the outer patch in the $\epsilon\to0$ limit).

On a chamber $C$, write
$\boldsymbol{\cF}=\boldsymbol b_C+k_C\boldsymbol{x}$.
If $\boldsymbol{x}_0\in C$ is a zero, then
$\boldsymbol{\cF}(\boldsymbol{x})=k_C(\boldsymbol{x}-\boldsymbol{x}_0)$, so
\begin{equation}
 Z_C=C\cap(\boldsymbol{x}_0+\ker k_C),\qquad
 \dim Z_C=r_G-\operatorname{rank}k_C.
 \label{eq:chambergeometry}
\end{equation}
In particular, an isolated interior
zero has invertible $k_C$.

Thus the continued zero set is assembled from affine pieces clipped by chamber
inequalities, including their boundary limits. These can be points, line segments, polygons, or higher-dimensional hyperplanes limited by the dimension of the maximal torus. Here, we use numerical sampling to locate candidate hyperplanes in the holonomy space as an alternative to a computationally costly (beginning at rank $r_G \geq 3$) exact solution to a set of linear equations on all chambers.

Our proposed pipeline begins with a grid search via numerical sampling to obtain a noisy candidate zero set. Then, we perform a proximity clustering via Mathematica's \textit{density based spatial clustering of applications (with) noise} ``DBSCAN" to capture sets of connected lower dimensional hyperplanes and remove isolated zeroes. With a connected component of the zero set in hand as such, we then utilize \textit{local principal component analysis} (LPCA) denoising to flatten noise directions. Finally, we utilize a \textit{random sample consensus} (RANSAC) hyperplane-fitting algorithm to determine the hyperplanes inside each connected component of the zero set.\footnote{For further reference on these tools from machine learning and image processing, see \cite{raeisi2023machine,vincent2001detecting,schnabel2007efficient}.} One may then go solve the exact linear equations along a particular chosen hyperplane for a direct verification.



\subsection{Generating a candidate dataset}\label{eq:vacGen}

We sample independent holonomy coordinates on a grid of spacing
$a$ and retain points with $\bar V_D<\mathcal E_0$, with the numerical threshold $\mathcal E_0$ chosen to be much smaller than $a$. If $K_{\max}=\max_C\|k_C\|_2$, then because $\boldsymbol{\cF}$ is both affine and continuous, it is therefore Lipschitz continuous
\begin{equation}
 |\boldsymbol{\cF}(\boldsymbol{x})-
 \boldsymbol{\cF}(\boldsymbol{y})|
 \leq K_{\max}|\boldsymbol{x}-\boldsymbol{y}|.
 \label{eq:gridbound}
\end{equation}
On the full periodic grid, any zero is guaranteed to be within the distance $\sqrt{r_G}a/2$ of a grid point. We may then take 
\begin{equation}
    \mathcal E_0 > \frac{K_{\max}^2r_G}{4}a^2
\end{equation}
to ensure that any neighboring point passes the threshold. A conservative bound is
\begin{equation}
    \label{eq:lipschitzbound}
    K_{\max}\leq\frac12\sum_\rho|\rho|^2
\end{equation}
which grows as $r_G$ and the complexity of your theory. A conservative threshold as such prevents perturbative vacua from ``falling through'' the grid and ensures detection near an off-grid zero.

The number of grid points increases exponentially with the rank. To reduce the computational cost of the sampling one can restrict the search to a specific Weyl alcove, or use an adaptive refinement around detected zeroes with a large initial threshold, though neither approach was used in this work. 

\subsection{DBSCAN clustering and LPCA denoising}\label{subsec:DBSCAN/LPCA}

After generating a candidate zero set $Z_C$, we use a $\texttt{minpts} > 1$ DBSCAN search to cluster nearby zeroes, with a neighborhood radius exceeding $\delta \gtrsim a\sqrt{r_G}$ to catch diagonally-connected components. One must dynamically adjust the threshold and grid spacing to not over or undersample points. The former may merge pieces, while the latter may disconnect pieces.\footnote{We also note that this is an unquotiented domain. The holonomy space is the unit torus, and this can be implemented via the identification $d_T(\boldsymbol{x},\boldsymbol{y})=
\min_{\boldsymbol n\in\mathbb Z^{r_G}}
|\boldsymbol{x}-\boldsymbol{y}+\boldsymbol n|$}
The threshold $\bar V_D<\mathcal E_0$ thickens the exact zero set.
To see by how much, let $\boldsymbol{x}_0$ be a zero in chamber $C$
and let $\boldsymbol{u}$ be a unit eigenvector of $k_C$ with eigenvalue
$\lambda$. As long as $\boldsymbol{x}_0+t\boldsymbol{u}$ remains in $C$,
\[
 \bar V_D(\boldsymbol{x}_0+t\boldsymbol{u})
 = |t k_C\boldsymbol{u}|^2 = \lambda^2 t^2 .
\]
Thus $\lambda=0$ gives a direction tangent to the zero set.
For $\lambda\ne0$, the cut admits displacements
$|t|<\sqrt{\mathcal E_0}/|\lambda|$ in that transverse direction.
These half-widths can vary from chamber to chamber.

LPCA estimates an affine tangent space from the covariance
matrix of neighboring points, retains directions with large
eigenvalues, and projects onto the corresponding affine space
through the local mean. Our method uses the $2r_G$ nearest neighbors to cover the nearest grid points, and a variance cutoff $\texttt{varTh}=(ha)^2$ to collapse any thickness smaller than $h$ grid units. It would be natural to take
\begin{equation}
    h\gtrsim\frac{\mathcal{E}_0}{(|\partial_j\bar{V}_D|_\text{min}\cdot a)}
\end{equation}
with the minimum taken over the nonzero values of $|\partial_j\bar{V}_D|$ in the cluster, since noise arises from low (but nonzero) variations of $\bar{V}_D$ near actual zeros. Additionally, near intersections, endpoints, and closely spaced cells, LPCA can move points between components  or erase genuine directions. Denoising is therefore used only to propose fits. Dimensions should be checked against $\ker k_C$, and fitted points against
the original coefficient and chamber inequalities.

\subsection{RANSAC hyperplane detection}

The final step uses RANSAC as a hyperplane-fitting algorithm to
iteratively identify densely clustered subsets of points and remove
their inliers. It can detect lines ($d_{\text{flat}}=1$) and planes
($d_{\text{flat}}=2$), though not simultaneously. In the examples
treated below, we choose $d_{\text{flat}}$ at each step to specify
the dimension of the object sought. More generally, a dynamic variant
could start with the largest candidate dimension and decrease
$d_{\text{flat}}$ once no further piece of that dimension is detected.

Points are labeled ``inliers'' if their distance from the fitted
affine piece falls below a threshold, which we set to
$1.5\hat\sigma$, where $\hat\sigma$ is the estimated noise level in
the data set. To estimate this noise level, we take the $k$ nearest
neighbors of each point, form the matrix of summed squared
deviations, and compute its eigenvalues in descending order.
The $(d_{\text{flat}}+1)$-th eigenvalue, denoted $\lambda_{d_{\text{flat}}+1}$, measures noise in a transverse direction.
For lines in three dimensions, there are two transverse directions.
Since our data have the same noise distribution in each direction,
we use the larger transverse eigenvalue to estimate the noise.
The eigenvalue $\lambda_{d_{\text{flat}}+1}$ measures the sum
of squared deviations along the leading transverse principal
direction. Its square root therefore provides a distance scale
that grows with the transverse thickness of the point cloud.
We use the empirical prescription
\begin{equation}
    \hat\sigma =
    \sqrt{\frac{\lambda_{d_{\text{flat}}+1}}{8d_{\text{flat}}}}.
\end{equation}
The denominator sets how restrictively this scale is used:
at fixed transverse scatter, the factor $d_{\text{flat}}$
makes the tolerance smaller for higher-dimensional fits,
while $8$ supplies an additional numerical adjustment.
These factors are tuning choices rather than a statistical
normalization of the noise variance.

We allocate $N_t=150$ trials to each detection. To assess this
choice, suppose the target component contains a fraction $w$
of the current cluster and each trial samples $m$ points.
The probability that all sampled points belong to that component
is approximately $w^m$. Assuming independent trials and
nondegenerate samples, the probability of obtaining at least
one such sample is therefore \cite{schnabel2007efficient}
\begin{equation}
 p=1-(1-w^m)^{150}.
 \label{eq:RANSACtrials}
\end{equation}
For two-point line fits, $m=2$. Assuming approximately equal
populations among the candidate lines, the six-line example
has $w\simeq1/6$, giving
\[
 p\simeq1-\left(\frac{35}{36}\right)^{150}\simeq0.985.
\]
Thus 150 trials meet the nominal 98\% target in this case.
As detected lines are removed, the remaining lines occupy
larger fractions of the data, improving this estimate.
Smaller component fractions or degenerate samples can require
more trials.

Repeating the 150-trial search for each of the
$N_{\mathrm{detections}}$ expected fits gives a total budget
of $150\,N_{\mathrm{detections}}$ trials per run.

Sequential fitting can reveal several affine pieces in one connected
cluster, but removing inliers can obscure shared intersections.
Disconnected coplanar cells can also be assigned to one fit.
The ideal higher-rank output is therefore a list of candidate affine
pieces and their support, followed by checks against
Eq.~\eqref{eq:chambergeometry} and the chamber boundaries.
For rational input data, these checks can be performed with
exact arithmetic.

The rank-two and rank-three examples below illustrate this method.
In the rank-two $(A_1,A_4)$ case, we also use exact linear analysis
to certify that all perturbative vacua have been captured; see the
ancillary Mathematica notebook.



\section{Applications: twist dependence and higher-rank geometry}
\label{sec:examples}

The applications in this section complement the rank-one examples of the companion
Letter~\cite{CompanionPRL}.
We first consider how the perturbative Coulomb locus
changes as flavor twists are varied in $\mathrm{SU}(2)$
$\mathcal{N}=4$ super-Yang--Mills theory.
We then study higher-rank $U(1)_r$-twisted
Argyres--Douglas reductions, where extended zero loci
motivate the reconstruction methods of
Section~\ref{sec:ML}.

The tree-level FI parameters vanish in these applications.
By \eqref{eq:QvsF}, the D-term equations can therefore
be written as $\nabla Q=0$, and the normalized diagnostic
is $\bar V_D=|\nabla Q|^2$.
We use this description to investigate the geometry
of the coefficient's zero set. Its interpretation as
a perturbative vacuum locus is subject to the outer/inner-patch
distinction explained above.

\subsection{$\mathrm{SU}(2)$ $\mathcal{N}=4$ SYM with flavor-twisted boundary conditions}\label{subsec:N=4ex}

Our first application is $\mathrm{SU}(2)$
$\mathcal{N}=4$ super-Yang--Mills theory, used here to
study the dependence of the perturbative Coulomb locus
on continuously variable flavor twists. From an $\mathcal{N}=1$ perspective, the $\mathcal{N}=4$ theory has SU(3) flavor symmetry. This group has rank two, so we can associate two chemical potentials $\xi_{1,2}$ to it, which we take to be real-valued for simplicity. We can alternatively consider three real-valued $\xi_j$ subject to $\xi_1+\xi_2+\xi_3\in\mathbb{Z}$. These correspond to imposing flavor-twisted boundary conditions on the three $\mathcal{N}=1$ chiral multiplets in the $\mathcal{N}=4$ theory: the twist phase for the $j$'th chiral multiplet as it goes around $S^1$ is set to $e^{2\pi i\xi_j}.$ Note that this definition makes $\xi_j$ 1-periodic.

\begin{figure}[t]
\centering
    \includegraphics[scale=.6]{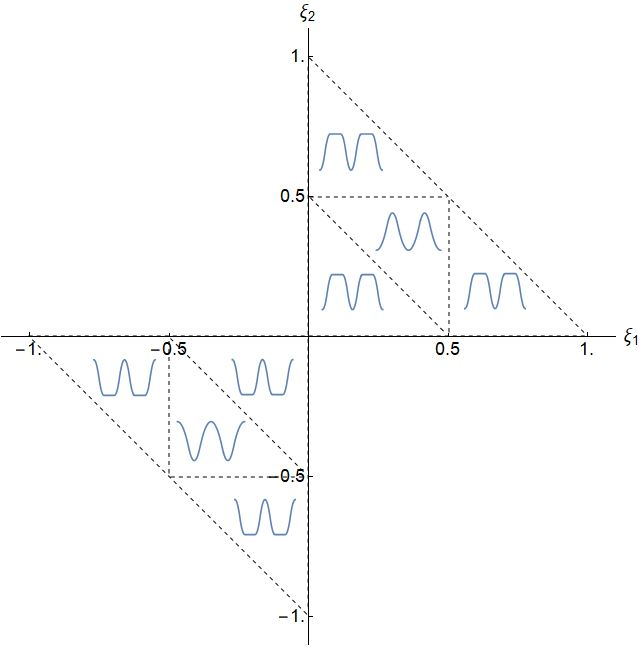}
\caption{The qualitative profile of the function $Q(x;\xi_{1,2})$, drawn over the range $x\in(-1,1)$, for fixed
$\xi_{1,2}$, in the two
complementary wings of the
space of the control-parameters $\xi_{1,2}$. (Note that $\xi_{1,2}$ are 1-periodic: $\xi_{1,2}\sim \xi_{1,2}+1$, so the two wings cover a fundamental domain.) The dashed lines indicate the bifurcation sets across which the qualitative behavior changes. On the outer patch, the leading potential is proportional to ${Q'}^2$, so intervals with $Q'=0$ give zero-potential directions. Threshold points require inner matching. The plot is taken from \cite{Ardehali:2021irq}.
\label{fig:CatastSimp}}
\end{figure}

The $Q$ function can be easily computed (cf. $V_2^\text{out}$ in \cite{Ardehali:2021irq}):
\begin{equation}
    Q(x;\xi_{1,2})=\frac{1}{3!}\, \sum_{a=1}^3\big[\big(\overline{B}_3(\xi_a+2x)+\overline{B}_3(\xi_a-2x)\big)+\overline{B}_3(\xi_a)\big].
\end{equation}
The qualitative profile of this $Q$ function is shown in
Figure~\ref{fig:CatastSimp}, adapted from Ref.~\cite{Ardehali:2021irq}.
Its bifurcations organize the changes of the continued stationary
locus as the twists vary. An isolated off-wall stationary point
has $Q''\ne0$ and gives a topological gauge sector; an interval
with $Q'=0$ gives a perturbative Coulomb direction. Wall stationary
points require the matched inner theory. The profile indicates that gapless vacua arise outside the inner triangles on each wing. At special twists
neutral chiral modes can also remain massless and must be included
when discussing the full spectrum.

\subsection{Higher-rank $U(1)_r$-twisted Argyres--Douglas theories}
\label{sec:AGtheories}

Our higher-rank applications use the Maruyoshi--Song
descriptions of the $(A_1,A_4)$, $(A_1,A_6)$, and $(A_1,A_7)$
Argyres--Douglas theories
\cite{Maruyoshi:2016aim,Agarwal:2016pjo},
with the R-twisted boundary condition relevant to
the SCFT/VOA correspondence
\cite{Beem:2013sza,Dedushenko:2023cvd}.
The rank-two example is supplemented with an exact chamberwise analysis;
the rank-three examples illustrate numerical reconstruction
of extended zero loci.

The supersymmetric $U(1)_r$ twist is
\begin{equation}
 \varphi(y+2\pi R_1)=e^{i\pi\gamma(\mathsf F+2r^{}_{\mathcal{N}=2})}\varphi(y),
 \qquad \gamma\in\mathbb Z,
 \label{eq:rTwist}
\end{equation}
where $\mathsf F$ is fermion number and $r^{}_{\mathcal{N}=2}$ has the
$\mathcal N=2$ normalization of Refs.~\cite{Dedushenko:2023cvd,ArabiArdehali:2024ysy}.
The compensating fermion-number factor preserves four supercharges.
On scalar components the twist shift is $\gamma \,r_{\mathcal{N}=2}^\chi$;
we take $\gamma=1$. Its period depends on the charge lattice.
The scalar KK formula therefore follows by replacing
$q^\chi\cdot\boldsymbol\xi$ with $\gamma \,r_{\mathcal{N}=2}^\chi$.

\subsubsection{$(A_1,A_4)$}\label{subsubsec:A1A4}

The Maruyoshi-Song description of the $(A_1,A_4)$ theory has gauge group $\mathrm{Sp}(2)$
(rank two, with a four-dimensional fundamental), with roots $\pm2e_j$ and
$\pm e_1\pm e_2$. The adjoint has shift $1/7$ and the two fundamentals have
shifts $3/7$ and $-2/7$. Up to an additive constant,
\begin{equation}
\begin{split}
6\,Q(x_1,x_2)
 &=\sum_{j=1}^{2}\Big[\overline B_3\big(2x_j+\tfrac17\big)
   +\overline B_3\big(-2x_j+\tfrac17\big)
   +\overline B_3\big(x_j+\tfrac37\big)+\overline B_3\big(-x_j+\tfrac37\big)\\
   &\qquad\qquad
   +\overline B_3\big(x_j-\tfrac27\big)+\overline B_3\big(-x_j-\tfrac27\big)\Big]\\
 &\hspace{-1.5cm}+\overline B_3\big(x_1+x_2+\tfrac17\big)+\overline B_3\big(x_1-x_2+\tfrac17\big)
   +\overline B_3\big(-x_1+x_2+\tfrac17\big)+\overline B_3\big(-x_1-x_2+\tfrac17\big).
\end{split}
\label{eq:QA1A4}
\end{equation}
The holonomies are $1$-periodic, and the Weyl alcove is
$0\le x_2\le x_1\le\tfrac12$.

\begin{figure}[t]
\centering
\includegraphics[width=0.6\linewidth]{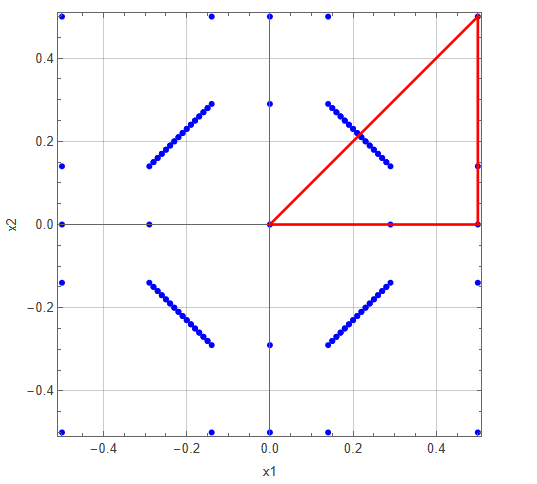}
\caption{Approximate zeros of $\bar V_D$ for the $(A_1,A_4)$ theory in the
$(x_1,x_2)$ plane. The Weyl alcove $0\le x_2\le x_1\le\tfrac12$ is highlighted
in red. Inside it there are five isolated points and a segment on the line
$x_1+x_2=\tfrac37$, which ends on the wall $x_1=x_2$.}
\label{fig:a1a4alcove}
\end{figure}

Write the sum in \eqref{eq:QA1A4} as a sum over chiral weights $\chi$ with
charge vectors $\rho^\chi$ and shifts $r_{\mathcal{N}=2}^\chi$. Then
\begin{equation}
\partial_iQ=\tfrac12\sum_\chi \rho^\chi_i\,\overline B_2\big(\rho^\chi\!\cdot x+r_{\mathcal{N}=2}^\chi\big),
\qquad
(k_C)_{ij}=-\partial_i\partial_jQ=-\sum_\chi \rho^\chi_i \rho^\chi_j\,
\overline B_1\big(\rho^\chi\!\cdot x+r_{\mathcal{N}=2}^\chi\big),
\label{eq:A1A4kC}
\end{equation}
where the second-derivative identity holds within chambers.
At a wall, the symmetric prescription $\overline B_1(n)=0$
for $n\in\mathbb Z$ gives the matched Cartan CS matrix after
subtracting the massless chirals. Here $k_C$ uses the sign
convention of Eq.~\eqref{eq:kout}. Since the weights come in
pairs $\pm \rho^\chi$ with equal shifts, the quadratic part of $\partial_iQ$
cancels. Hence $\nabla Q$ is continuous and piecewise affine, with Jacobian
$-k_C$ in each chamber. The solution set of $\nabla Q=0$ in a chamber is
then an affine subspace of dimension $\dim\ker k_C$, which is the dimension
check of Section~\ref{sec:ML}. Solving chamber by chamber in exact rational
arithmetic inside the alcove gives exactly five isolated zeros and one
segment, in agreement with the sampled locus of
Figure~\ref{fig:a1a4alcove}:
\begin{equation}
\begin{aligned}
&(0,0),\quad(\tfrac27,0),\quad(\tfrac12,0),\quad(\tfrac12,\tfrac17),\quad(\tfrac12,\tfrac12),\\
&\mathcal I=\big\{x_1+x_2=\tfrac37,\ \tfrac3{14}\le x_1\le\tfrac27\big\}.
\end{aligned}
\label{eq:A1A4zeroset}
\end{equation}
The remaining points in Figure~\ref{fig:a1a4alcove} are images of these under
the Weyl group and integer shifts, and have identical physics.

\paragraph{Isolated zeros.}
All five isolated zeros lie on walls. Matching gives the
3d $\mathcal N=2$ gauge theories listed in
Table~\ref{tab:A1A4points}, whose Cartan CS matrices are
nondegenerate. Each simple factor $G$ carries an $\mathcal N=2$
Chern--Simons level $K$. For $\mathrm{SU}(2)_{(j)}$ generated by the root
$2e_j$, the fundamental has charges $\pm1$ under the Cartan $U(1)_j$, so
$K=\tfrac12(k_C)_{jj}$. For $\mathrm{Sp}(2)$, $k_C=2K\,\mathbbm{1}$. As a check,
a positive-real-mass adjoint chiral shifts $K$ by $h^\vee$.

\begin{table}[t]
\centering
\renewcommand{\arraystretch}{1.3}
\begin{tabular}{@{}cccll@{}}
\toprule
$(x_1,x_2)$ & $k_C$ & unbroken gauge group & light chirals & non-abelian level \\
\midrule
$(0,0)$ & $\mathrm{diag}(4,4)$ & $\mathrm{Sp}(2)$ & none & $\mathrm{Sp}(2)_{2}$ \\
$(\tfrac27,0)$ & $\mathrm{diag}(-\tfrac32,2)$ & $U(1)_1\times\mathrm{SU}(2)_{(2)}$
  & one, $q=+1$ under $U(1)_1$ & $\mathrm{SU}(2)_{1}$ \\
$(\tfrac12,0)$ & $\mathrm{diag}(2,2)$ & $\mathrm{SU}(2)_{(1)}\times\mathrm{SU}(2)_{(2)}$
  & none & $\mathrm{SU}(2)_{1}\times\mathrm{SU}(2)_{1}$ \\
$(\tfrac12,\tfrac17)$ & $\mathrm{diag}(2,-2)$ & $\mathrm{SU}(2)_{(1)}\times U(1)_2$
  & none & $\mathrm{SU}(2)_{1}$ \\
$(\tfrac12,\tfrac12)$ & $\mathrm{diag}(4,4)$ & $\mathrm{Sp}(2)$ & none & $\mathrm{Sp}(2)_{2}$ \\
\bottomrule
\end{tabular}
\caption{Data at the isolated zeros of $\nabla Q$ in the alcove. The
off-diagonal entry of $k_C$ vanishes at all five points. At $(\tfrac27,0)$ the
massless chiral comes from the fundamental with shift $-\tfrac27$ and is a
singlet of $\mathrm{SU}(2)_{(2)}$. Its presence makes $(k_C)_{11}$
half-integral, as required by the parity anomaly.}
\label{tab:A1A4points}
\end{table}

To complete the picture, we now add known infrared information,
beyond the perturbative matching above, as in the examples of
the accompanying Letter \cite{CompanionPRL}.
Pure $\mathcal N=2$ Yang--Mills--Chern--Simons theory with simple
group $G$ has supersymmetric vacua for $|K|\ge h^\vee$, with
bosonic level $K-h^\vee\,\mathrm{sgn}\,K$ after integrating out
the adjoint gaugino. For $0<|K|<h^\vee$ supersymmetry is spontaneously
broken \cite{Witten:1999ds,Ohta:1999iv,Bergman:1999na,Intriligator:2013lca}.
Here $h^\vee=3$ for $\mathrm{Sp}(2)$ and $h^\vee=2$ for $\mathrm{SU}(2)$.
Table~\ref{tab:A1A4points} gives $|K|=2<3$ at $(0,0)$ and $(\tfrac12,\tfrac12)$,
and $|K|=1<2$ for every $\mathrm{SU}(2)$ factor at the other three points. At
$(\tfrac27,0)$ and $(\tfrac12,\tfrac17)$ the abelian sector factorizes from the
$\mathrm{SU}(2)$ factor: the light chiral at the former point is neutral
under $\mathrm{SU}(2)$, and the latter has no light chirals.
The pure nonabelian factor therefore breaks supersymmetry in
each case, irrespective of the abelian sector and the sign of $K$.
Hence none of the five isolated zeros
supports a supersymmetric vacuum, even though $\nabla Q=0$ holds there.

\paragraph{The interval.}
On $\mathcal I$ the CS matrix is constant in the interior,
\begin{equation}
k_C\big|_{\mathcal I}=-\begin{pmatrix}1&1\\1&1\end{pmatrix},
\qquad \ker k_C=\mathrm{span}\{(1,-1)\},
\end{equation}
and the kernel is the tangent direction of $\mathcal I$. In the interior the
gauge group is broken to $U(1)^2$ and no chiral multiplet is massless. Only
the photon combination $A_1+A_2$ has a Chern--Simons term, while $A_1-A_2$ has
none. The segment is therefore a perturbatively flat Coulomb direction, not a
gapped topological sector. Its two ends are distinct:
\begin{itemize}
\item At $(\tfrac3{14},\tfrac3{14})$, on the wall $x_1=x_2$, the root
$e_1-e_2$ is massless. The coroot is $(1,-1)$, so the level of the enhanced
$\mathrm{SU}(2)$ is $K=\tfrac12(1,-1)\,k_C\,(1,-1)^T=0$, and no light matter
is charged under it. The interval is the perturbative Coulomb branch of this
$\mathrm{SU}(2)_0$ factor, and the perturbative analysis stops here.
\item At $(\tfrac27,\tfrac17)$ two chirals become massless simultaneously, with
charges $(-1,1)$ (from the adjoint) and $(1,0)$ (from the fundamental with shift
$-\tfrac27$). Across this threshold $\nabla Q\neq0$, so the zero set ends.
\end{itemize}

To summarize, within the Weyl alcove, the continued-coefficient equation $\nabla Q=0$
gives five isolated points and the interval $\mathcal I$. The
supplemental infrared input excludes supersymmetric vacua at
all five isolated points. Therefore among these coefficient zeros, only
$\mathcal I$ remains a candidate locus. Whether any point of $\mathcal I$ survives is not
decided by the outer-patch data above. It requires inclusion of nonperturbative
(monopole-instanton) superpotential terms, and a separate analysis of the matched inner theory at
the wall endpoint and at the threshold. This is beyond the scope of the
present analysis; see \cite{ArabiArdehali:2024ysy} for the nonperturbative analysis indicating that only the point $(\tfrac27,\tfrac17)$ on the interval survives, supporting a SUSY-enhancing Gang-Kim-Stubbs theory \cite{Gang:2023rei} in the infrared.

\subsubsection{$(A_1,A_6)$}\label{subsubsec:A1A6}

The $(A_1,A_6)$ Lagrangian has gauge group $\mathrm{Sp}(3)$,
with $\mathrm{Sp}(1)=\mathrm{SU}(2)$. Its three independent
coordinates have roots $\pm2e_j$ and $\pm e_i\pm e_j$. We follow the convention of \cite{ArabiArdehali:2024vli} for the R-symmetry shifts per chiral multiplet, as well as for constant terms that do not contribute to the holonomy saddle, which are nonetheless given for a complete description of the $Q(x_j)$ function.

In terms of the parameters $\alpha_j = \frac{2(4+j)}{9},\ \beta_j = \frac{2j}{9},$
the $Q$ function of the R-twisted $(A_1, A_6)$ comes out (cf.~\cite{ArabiArdehali:2024ysy, ArabiArdehali:2024vli})
{\small\begin{equation}
\begin{split}
    6Q(x_j) &= 3\overline{B}_3\big(\frac{1}{9}\big) + \sum_{j = 1}^3 \left(\overline{B}_3(\alpha_j) - \overline{B}_3(\beta_j) + \overline{B}_3\big(2x_j + \frac{1}{9}\big) + \overline{B}_3\big(-2x_j + \frac{1}{9}\big)\right)\\ 
    &+ \sum_{j = 1}^3 \left(\overline{B}_3\big(x_j + \frac{4}{9}\big) + \overline{B}_3\big(-x_j + \frac{4}{9}\big) + \overline{B}_3\big(x_j - \frac{3}{9}\big) + \overline{B}_3\big(-x_j - \frac{3}{9}\big)\right)\\ 
    &\hspace{-2.3cm}+ \sum_{1 \leq i < j \leq 3}\!\! 
    \bigg(\overline{B}_3\big(x_i + x_j + \frac{1}{9}\big) + \overline{B}_3\big(x_i - x_j + \frac{1}{9}\big)+ \overline{B}_3\big(-x_i + x_j + \frac{1}{9}\big) + \overline{B}_3\big(-x_i - x_j + \frac{1}{9}\big)\bigg).
    \end{split}
\end{equation}}

We use the algorithm of Section~\ref{sec:ML} to explore the stationary locus in the unit cube. The fitted lines and planes are candidates for chamberwise affine pieces, whose physical interpretation requires the patch and symmetry checks described there.

First, we generate a candidate zero dataset as follows. The exploratory computation estimates a maximum of order $10$ for $\bar V_D$ on the unit cube and divides by that estimate. We then subject the normalized $\bar{V}_D$ to a zero search, to within a small threshold of order $10^{-3}$, on a grid of step size $a=0.01$.

\begin{figure}[h!]
    \centering
    \includegraphics[width=0.6\linewidth]{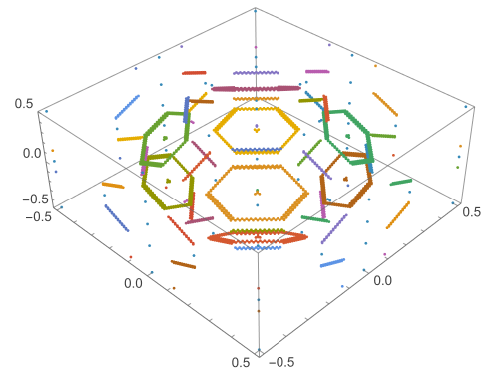}
    \caption{DBSCAN of the approximate zeros of $\bar V_D$ for $(A_1, A_6)$ in the unit cube. The sampled locus suggests segments and isolated points. Their completeness, wall assignment, and identifications require separate checks.}
    \label{fig:a1a6clusters}
\end{figure}

Next, we cluster the approximate zeros. See Figure~\ref{fig:a1a6clusters} for the approximate zeroes after DBSCAN clustering as described in Section~\ref{sec:ML}. %

\begin{figure}[h!]
    \centering
    \includegraphics[width=0.8\linewidth]{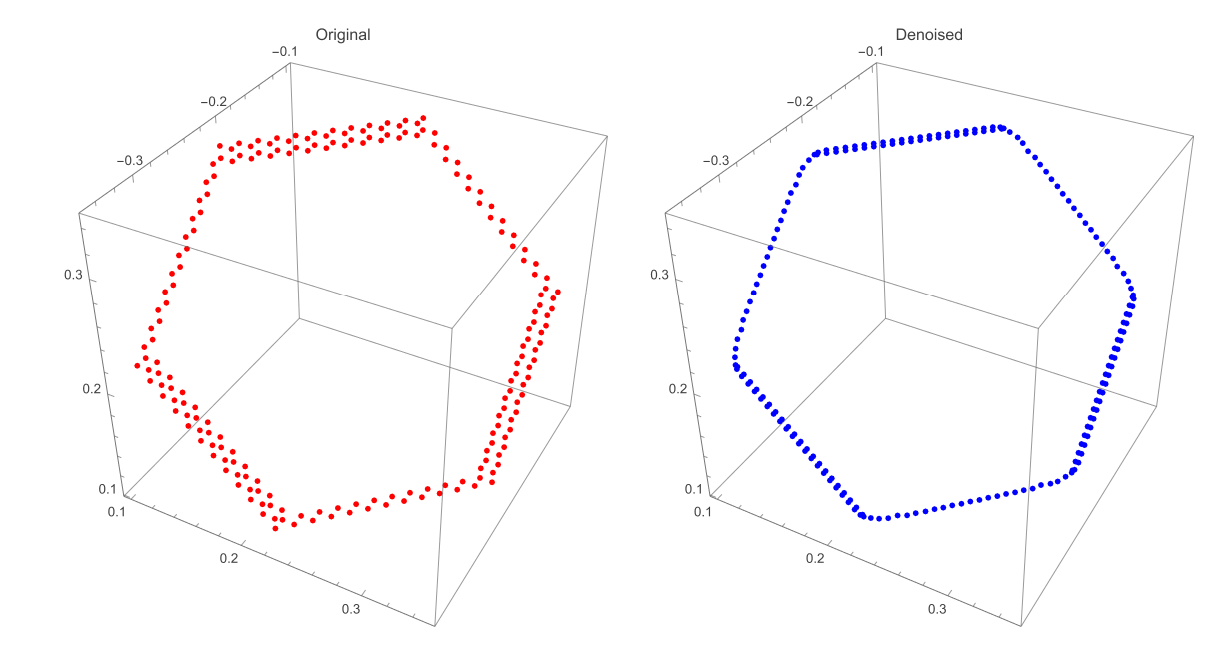}
    \caption{LPCA applied to a candidate hexagonal cluster of the
$(A_1,A_6)$ coefficient zero set. Projection reduces transverse
thickness but moves the sampled points.}

    \label{fig:denoisedhexagon}
\end{figure}

The threshold gives each candidate component a finite thickness.
We denoise our clusters using LPCA. This is illustrated on a ``hexagon cluster" in Figure~\ref{fig:denoisedhexagon}, where LPCA reduces the transverse thickness. This visual simplification does not itself establish the dimension or preserve intersections.%

\begin{figure}[h!]
    \centering
    \includegraphics[width=0.4\linewidth]{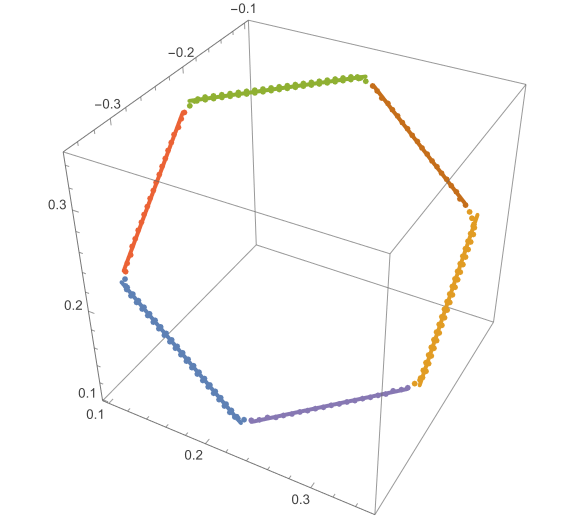}
    \caption{One RANSAC line fit in a denoised hexagonal cluster for $(A_1,A_6)$. The overlay illustrates the fit to the projected sample.}
    \label{fig:linedetection}
\end{figure}

Once the cluster has been denoised, it is ready for RANSAC-based hyperplane detection, which in the case of the hexagon cluster returns six candidate line components, as in Figure~\ref{fig:linedetection}. %

These examples illustrate the complementary roles of exact
chamberwise analysis and numerical reconstruction of the
perturbative coefficient zeros. The matched theories provide
the starting point for further applications to the SCFT/VOA correspondence
\cite{Beem:2013sza,ArabiArdehali:2024ysy,ArabiArdehali:2024vli}. We leave out a detailed analysis, as in the $(A_1,A_4)$ case above, showing that all the perturbative vacua are lifted non-perturbatively, leaving only some interval endpoint supporting a 3d SUSY-enhancing Gang-Kim-Stubbs theory as in \cite{ArabiArdehali:2024vli}.

\subsubsection{$(A_1,A_7)$}\label{subsubsec:A1A7}

The $(A_1,A_7)$ Lagrangian has gauge group $\mathrm{SU}(4)$, which, like the
$\mathrm{Sp}(3)$ of the $(A_1,A_6)$ case, has rank three, so the same
search in the unit cube applies. We take three independent coordinates
$x_1,x_2,x_3$ and set $x_4=-x_1-x_2-x_3$, so that the roots are $e_i-e_j$
with $i\neq j$.

The $Q$ function of the R-twisted $(A_1,A_7)$ then takes the form
(cf.~\cite{ArabiArdehali:2024ysy})
{\small\begin{equation}
\begin{split}
    6Q(x_j) &= 3\overline{B}_3\big(\frac{1}{10}\big) + \sum_{1 \leq i < j \leq 4}
    \Big(\overline{B}_3\big(x_i - x_j + \frac{1}{5}\big)
    + \overline{B}_3\big(-x_i + x_j + \frac{1}{5}\big)\Big)\\
    &\quad + \sum_{i = 1}^{4}
    \Big(\overline{B}_3(\alpha_j) + \overline{B}(\beta_j) + \overline{B}_3\big(x_i - \frac{3}{10}\big)
    + \overline{B}_3\big(-x_i - \frac{3}{10}\big)\Big).
\end{split}
\end{equation}}
Here, $\alpha_j = \frac{9-j}{5}$ and $\beta_j = \frac{j+1}{5}$ and the shifts are fixed by R-symmetry \cite{ArabiArdehali:2024vli}.

We apply the algorithm of Section~\ref{sec:ML} exactly as for $(A_1,A_6)$:
normalization of $\bar V_D$, a zero search to within the same threshold on the
same grid, DBSCAN clustering, LPCA denoising, and RANSAC-based flat detection.
Novel to this example is that the middle of the unit cube contains dimension-2 hyperplanes connected by dimension-1 hyperplanes.

\begin{figure}[h!]
    \centering
    \includegraphics[width=0.6\linewidth]{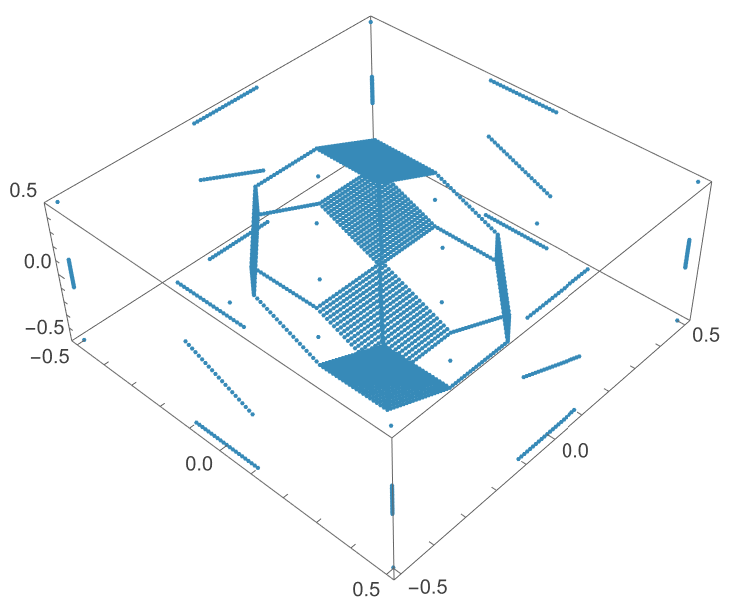}
    \caption{Approximate zeros of $\bar V_D$ for $(A_1, A_7)$ in the unit cube,
    before clustering. Besides isolated points and short segments, the central
    cluster contains thin edge-like strands and filled sheets. The
    sampled locus suggests components of dimension-1 and dimension-2.}
    \label{fig:a1a7unclustered}
\end{figure}

Figure~\ref{fig:a1a7unclustered} shows the approximate zeros in the unit cube
$[-\tfrac12,\tfrac12]^3$ before clustering. As in the $(A_1,A_6)$ case there
are isolated points and short segments near the boundary of the cube. The
central polyhedral cluster, however, contains filled sheets in addition to
one-dimensional edges.

\begin{figure}[h!]
    \centering
    \includegraphics[width=0.9\linewidth]{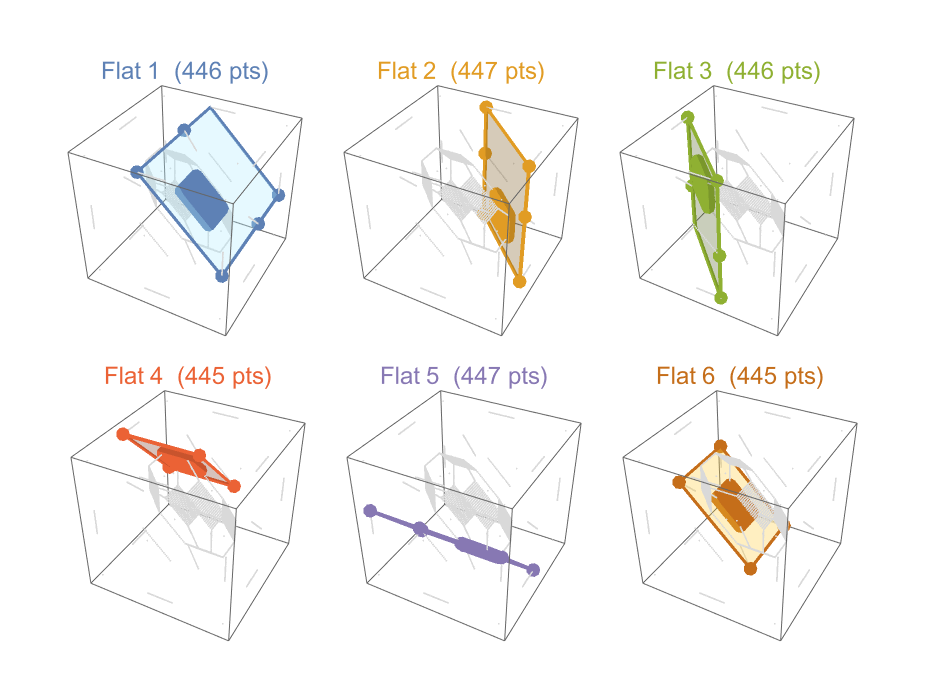}
    \caption{RANSAC flat detection on the denoised central cluster of the
    $(A_1,A_7)$ coefficient zero set. Each panel shows one candidate flat
    (inlier counts in the panel titles), its inliers, and the convex hull of
    the inliers within the fitted plane; the remaining points are shown in
    gray.}
    \label{fig:a1a7planes}
\end{figure}

As before, the threshold gives each candidate component a finite thickness,
so the appearance of a filled sheet in Figure~\ref{fig:a1a7unclustered} does
not by itself establish that the underlying component is two-dimensional.
We therefore denoise the central cluster with LPCA and run RANSAC on the
result. In the $(A_1,A_6)$ example, the flats returned for the hexagon cluster
were lines. Here the procedure is not tied to line components, and it returns
six candidate flats for the central cluster of $(A_1,A_7)$, shown in
Figure~\ref{fig:a1a7planes}. The fitted objects are planes in the cube, that is,
two-dimensional affine pieces of the zero locus. Some of them appear nearly
edge-on from the chosen viewpoint.

Compared with $(A_1,A_6)$, where the reconstructed components were lines, the
$(A_1,A_7)$ example shows that the same numerical procedure also resolves
planes in the moduli space. The method therefore does not assume the
dimension of the components in advance, and applies to loci that mix points,
lines, and planes within a single cube.

As in the $(A_1,A_6)$ case, we leave out a detailed analysis showing that all the perturbative vacua are lifted non-perturbatively, leaving only a junction point supporting a 3d SUSY-enhancing theory as in \cite{ArabiArdehali:2024vli}.

\section{Discussion and future directions}\label{sec:discussion}

The perturbative matching problem has a simple global answer:
the periodic Bernoulli sum determines the D-linear coefficient,
while subtraction of retained modes determines the FI and CS
data on every inner patch (as captured globally via the modified Bernoulli replacements). Uniform control of the leading
outer potential requires both a regular kinetic metric and
a valid expansion in the auxiliary fields. With these conditions
in place, outer zeros are found from \eqref{eq:vacEqs} without computing the
full kinetic metric.

The companion Letter emphasizes the resulting vacuum sectors.
Here the affine chamber structure makes their higher-rank
analysis a problem in linear equations and inequalities,
with numerical reconstruction useful for exploration.

Determining the infrared vacuum structure of massless interacting
sectors remains a separate dynamical problem. The supplemental
results in the $(A_1,A_4)$ example illustrate how independent
infrared information can complete the picture obtained from matching.

Nonperturbative F-terms pose a further dynamical question.
Gauge and global charges constrain possible monopole
superpotentials, and there are globally valid formulae for such charges that can help determine naturally generated F-terms \cite{ArabiArdehali:2024ysy}. Zero modes and their interactions must
also be analyzed. A systematic study could extend the circle
reduction of chiral models such as those in
Refs.~\cite{Poppitz:1998vd,Amariti:2015kha,Aghaei:2017xqe,Razamat:2017hda,Lee:2019hzq,ArabiArdehali:2024vli}.
Such nonperturbative F-terms are the subject of a follow-up work.

\begin{acknowledgments}
    AA thanks Antonio Amariti, Cyril Closset, Thomas Grimm, Erich Poppitz, Sadegh Raeisi, and Andrea Zanetti for related discussions; INFN Milano, the University of Michigan in Ann Arbor, and the University of Toronto for hospitality; and the organizers of Aspects of Supersymmetric Quantum Field Theory at Seoul National University for the invitation to present some of these results. DJR would like to thank Martin Rocek and the Simons Center for Geometry and Physics  for hospitality during the workshop Symplectic Singularities, Supersymmetric QFT, and Geometric Representation Theory where some of this work was done. ChatGPT GPT-6-Astra helped in editing the draft, obtaining the control estimates in Section~\ref{sec:scalar_pot}, and drafting Appendix~\ref{app:KKren}. Claude Sonnet~5.5 helped in drafting Sections~\ref{subsubsec:A1A4} and \ref{subsubsec:A1A7}. GPT-6 and Sonnet~5.5 also helped with generating the code in the ancillary Mathematica notebook. The authors have thoroughly checked and revised the AI outputs.
\end{acknowledgments}

\appendix

\section{Compatibility with $S^3\times S^1$}\label{app:index}

This appendix develops the partition-function perspective
on the patchwise D-term equations.
We approach $\mathbb{R}^3\times S^1$ through the
large-$R_3$ limit of $S^3_{R_3}\times S^1$, and compare
the resulting stationary-phase equations with the
flat-space Wilsonian couplings.
The distinction between outer and inner patches is
essential: the former contain no light charged fields,
whereas the latter retain determinants associated
with the additional light multiplets.
The comparison identifies common coupling data and
candidate stationary loci, although it is worth emphasizing that it does not equate every
stationary point with a supersymmetric quantum vacuum.
The corresponding three-dimensional comparison,
between $\mathbb{R}^3$ and large-radius $S^3$,
is developed in Appendix~\ref{app:R3_from_S3}.

We assume a nonanomalous $U(1)_R$ symmetry and twists compatible
with the superpotential. Write $\beta=2\pi R_1/R_3$ and take
$\beta\to0^+$ at fixed $R_1$. The following asymptotics are
local on regions with fixed sets of light modes; global dominance
requires retaining the relative normalization of those regions.

The dimensionless 3d FI parameter is defined through  $\tilde{\zeta}:=R_3\zeta$. It is related to the dimensionless 4d FI parameter $\tilde{\zeta}^{(4)}:=(2\pi R_1)^2\,\zeta^{(4)}$ as in $\tilde{\zeta}=\frac{2\pi}{\beta}\tilde{\zeta}^{(4)}$. The 1-loop correction $\tilde{\zeta}^\text{1-loop}$ leads to 
\begin{equation}
    \tilde{\zeta}^\text{eff}:=\tilde{\zeta}+\tilde{\zeta}^\text{1-loop}=\frac{2\pi}{\beta}\tilde{\zeta}^{(4)}+\tilde{\zeta}^\text{1-loop}.\label{eq:zeta_tilde_eff}
\end{equation}

An FI insertion on compact $S^3\times S^1$ must respect
large gauge transformations \cite[Sec.~5]{Aharony:2013dha}.
In the convention below, the factor
$\exp[-i8\pi^3\tilde\zeta^{(4)}_jx_j/\beta^2]$
requires $-4\pi^2\tilde{\boldsymbol\zeta}^{(4)}/\beta^2$ to
pair integrally with the cocharacter lattice. In a unit-period
abelian coordinate its component is an integer. Continuous
flat-space FI inputs can be compared through admissible
large-radius sequences. We use $\mathcal I$ below for the resulting
supersymmetric partition function.

Assuming all chiral multiplet R-charges satisfy $0<r_\chi<2$, the localized expression reads\footnote{In comparing with Aharony et al \cite{Aharony:2013dha} note that $\zeta^{(4)}_\text{here}=-\xi^{(4)}_\text{there}$.} (up to an overall supersymmetric Casimir energy factor which is irrelevant to our discussion):
    \begin{equation}
	\mathcal{I}(q;\boldsymbol{v})=(q;q)^{2r^{}_{\!G}}\int_{\mathfrak{h}_{\text{cl}}} D\boldsymbol{x}\ e^{-i\frac{8\pi^3}{\beta^2} \tilde{\zeta}^{(4)}_jx_j}\ \frac{\prod_\chi\prod_{\rho}\Gamma_e\big(r_\chi {\tau}+\rho\cdot \boldsymbol{x}+q^\chi\cdot\boldsymbol{\xi}\big)}{\prod_{\alpha_+}\Gamma_e(\alpha_+\cdot\boldsymbol{x})\, \Gamma_e(-\alpha_+\cdot\boldsymbol{x})},\label{eq:EHIgen}
\end{equation}
Here $q=e^{-\beta}$, $\tau=i\beta/(2\pi)$,
$v_\ell=e^{2\pi i\xi_\ell}$, and
$\Gamma_e(z)=\Gamma(e^{2\pi iz};q,q)$ is the elliptic gamma
function in additive notation. The measure $D\boldsymbol{x}$
is the Cartan-torus measure divided by the Weyl order, with
equivalent normalization on a fundamental domain.
The symbol $\Delta_\chi$ below denotes the weight set of $\chi$.
Sending $R_3\to\infty$ at fixed $R_1$ is the Cardy limit
$\beta\to0^+$.

\subsection*{Outer patch}

On the outer patch, inside the integrand we use the simplification \cite{ArabiArdehali:2015ybk}
\begin{equation}
    \frac{\prod_\chi\prod_{\rho}\Gamma_e\big(r_\chi {\tau}+\rho\cdot \boldsymbol{x}+q^\chi\cdot\boldsymbol{\xi}\big)}{\prod_{\alpha_+}\Gamma_e(\alpha_+\cdot\boldsymbol{x})\, \Gamma_e(-\alpha_+\cdot\boldsymbol{x})}\xrightarrow{\beta\to0} e^{i\frac{8\pi^3}{\beta^2}Q(\boldsymbol{x})-\frac{4\pi^2}{\beta}L(\boldsymbol{x})},
\end{equation}
valid up to a locally $\boldsymbol{x}$-independent factor \cite{Ardehali:2021irq}. Gauge anomaly cancellation and a nonanomalous R-symmetry make the following functions piecewise quadratic and piecewise affine, respectively. The functions ${Q}(\boldsymbol{x})$ (resp. ${L}(\boldsymbol{x})$) are defined as \cite{ArabiArdehali:2015ybk}:
\begin{equation}
    \begin{split}
Q(\boldsymbol{x})&:=\frac{1}{12}\sum_{\chi}\sum_{\rho
\in\Delta_\chi}\kappa(\rho\cdot
\boldsymbol{x}+q^\chi\cdot\boldsymbol{\xi}),\\
L(\boldsymbol{x})&:= \frac{1}{2}\sum_{\chi}(1-r_\chi)\sum_{\rho
}\vartheta(\rho\cdot
\boldsymbol{x}+q^\chi\cdot\boldsymbol{\xi})-\sum_{\alpha_+}\vartheta(\alpha_+\cdot
\boldsymbol{x}),\label{eq:Q&L_Def}
    \end{split}
\end{equation}
where $\kappa(x)=2\overline{B}_3(x)$ and $\vartheta(x)=-\overline{B}_2(x)+1/6\,.$

The outer patch consists of multiple disjoint connected components. On each connected component, the functions ${Q},{L}$ are analytic, and admit Taylor expansion. Displaying the  $\boldsymbol{x}$-independent terms by const, we get from the expansions
\begin{equation}
    \begin{split}
        {Q}(\boldsymbol{x})&=-k^\text{1-loop}_{ij}\,\frac{x_i x_j}{2}-\frac{\beta}{2\pi}\,\tilde{\zeta}_j^\text{1-loop}x_j+\text{const}\,,\\
        {L}(\boldsymbol{x})&=-k^\text{1-loop}_{jR}x_j+\text{const}\,.
    \end{split}\label{eq:Qexp_Lexp}
\end{equation}
Here $k_{ij}^\text{1-loop},\tilde{\zeta}_{j}^\text{1-loop},k_{jR}^\text{1-loop}$ are the 1-loop corrections generated by integrating out the charged 3d multiplets which, by the definition of the outer patch, are all heavy there.

We thus find (up to nonzero overall factors constant on each connected component)%
\begin{equation}
\begin{split}
    \mathcal{I}_\text{outer patch}(q;\boldsymbol{v})&\xrightarrow{\beta\to0}
    \int_\text{outer patch} \!\!{\mathrm{d}^{r_G}x}
    \ \ e^{-i
    \frac{(2\pi)^3}{\beta^2}\,\text{\normalsize(} k^\text{eff}_{ij} \frac{x_i x_j}{2} \,+\,
    \frac{\beta}{2\pi}\tilde{\zeta}^\text{eff}_j x_j\text{\normalsize)}\,+\,\frac{4\pi^2}{\beta} \, k^\text{eff}_{jR}\, x_j },%
    \end{split}\label{eq:S3S1_Z_far_simp}
\end{equation}
with the effective couplings obtained as the sum of the classical and 1-loop pieces. Note that here 
\begin{equation}
    k^\text{eff}_{ij}=k^\text{1-loop}_{ij},\qquad k^\text{eff}_{jR}=k^\text{1-loop}_{jR},
\end{equation}
since there are no tree-level CS couplings descending from 4d. If similarly to \eqref{eq:AZ}:
\begin{equation}
    \boldsymbol{\xi}=\boldsymbol{\xi}^0+\frac{\beta}{2\pi}\tilde{\boldsymbol{m}}^{}_f\,,
\end{equation}
we package an $\tilde{\boldsymbol{m}}_f$ dependent contribution into $\tilde{\zeta}^\text{eff}_j$ as discussed below \eqref{eq:xi_plus_kjfmf}.

It will be important below that $\tilde{\zeta}_j^\text{eff}$ is of order $1/\beta$, and hence $\frac{\beta}{2\pi}\tilde{\zeta}_j^\text{eff}$ of order one, if either $\tilde{\zeta}_j^{(4)}$ is non-zero of order one (see \eqref{eq:zeta_tilde_eff}), or the constant piece of $\partial_j Q$ is non-zero of order one (see \eqref{eq:Qexp_Lexp}).

Within a chamber, the stationary equation for the leading oscillatory phase in \eqref{eq:S3S1_Z_far_simp} reads:
\begin{equation}
    0=
    \frac{2\pi}{\beta}\big(\tilde{\zeta}^{(4)}_j-\partial_{x_j}Q(\boldsymbol{x})\big)=k^\text{eff}_{ij} {\tilde{\sigma}_i} \,+\,
    \tilde{\zeta}^\text{eff}_j=: \widetilde F_j(\tilde{\boldsymbol{\sigma}})\,.\label{eq:F=0fromIndex}
\end{equation}
Here $\tilde{\boldsymbol\sigma}=R_3\boldsymbol\sigma=2\pi\boldsymbol x/\beta$ and $\widetilde F_j(\tilde{\boldsymbol\sigma})=R_3F_j(\tilde{\boldsymbol\sigma}/R_3)$; the tilde distinguishes this dimensionless quantity from the flat-space coefficient.


Extended saddles on the outer patch arise along some direction $\hat{\boldsymbol{g}}$ when the second derivative of the coefficient of $1/\beta^2$ in \eqref{eq:S3S1_Z_far_simp} satisfies:
\begin{equation}
    k^\text{eff}_{ij}\hat{{g}}_i=0,\qquad\text{for all $j$}.
\end{equation}
For magnetic charges in the allowed cocharacter lattice, the same kernel condition gives vanishing induced gauge charge of the abelian monopole operator,
\begin{equation}
    c_j(\boldsymbol{m})=-k^\text{eff}_{ij}m_i.
\end{equation}

Thus the first derivative $\partial_{{x}_j}$ of ${Q}$ encodes the dimensionless one-loop D-coefficient, while the second derivative $\partial_{x_i}\partial_{x_j}$ of ${Q}$ encodes the one-loop correction to $k_{ij}$ (and hence the gauge charge of BPS monopoles).

The $1/\beta^2$ phase therefore recovers the outer perturbative zero equations.

\subsection*{Inner patches}

Choose $\boldsymbol{x}^s$ in an inner patch and introduce
$\tilde{\boldsymbol\sigma}=2\pi(\boldsymbol{x}-\boldsymbol{x}^s)/\beta$.
Let $L_\chi$ and $L_V$ denote retained chiral weights and vector
roots, respectively. Write
$\mu_\rho^s=R_3M_\rho(\boldsymbol{x}^s)$ for a retained
dimensionless real mass, including any flavor deformation,
and define
\begin{equation}
 \tilde\zeta_{j,\mathrm{loc}}=R_3F^{\mathrm{in}}_j(\boldsymbol{x}^s/R_1).
 \label{eq:localFIindex}
\end{equation}
Up to locally constant factors, the inner contribution has
the matched form
\begin{equation}
\begin{split}
 \mathcal I_{\mathrm{in}}
 &\sim\int_{\frac{2\pi}{\beta}(\mathrm{in}-\boldsymbol{x}^s)}
 d^{r_G}\tilde\sigma\,
 e^{-2\pi i(\frac12k^{\mathrm{in}}_{ij}\tilde\sigma_i\tilde\sigma_j
 +\tilde\zeta_{j,\mathrm{loc}}\tilde\sigma_j)
 +2\pi k^{\mathrm{in}}_{jR}\tilde\sigma_j}\\
 &\quad\times
 \frac{\prod_{\rho\in L_\chi}
 \Gamma_h(i r_\chi+\rho\cdot\tilde{\boldsymbol\sigma}+\mu_\rho^s)}
 {\prod_{\alpha\in L_V}
 \Gamma_h(\alpha\cdot\tilde{\boldsymbol\sigma}+\mu_\alpha^s)}.
\end{split}
\label{eq:S3S1_Z_near}
\end{equation}
The vector product includes both signs of each retained root.
The convention is $\Gamma_h(z)=\Gamma_h(z;i,i)$ throughout
this appendix. Heavy modes determine the local FI and CS
couplings; the determinants of retained modes remain explicit.
For a mode massless at the reference wall, $\mu_\rho^s=0$
in the absence of a finite real-mass deformation. Retained
modes with nonzero reference masses require the displayed
offsets.

On a bounded region of $\tilde{\boldsymbol\sigma}$ with finite
$\mu_\rho^s$, the potentially large linear phase has coefficient
$\tilde\zeta_{\mathrm{loc}}=O(1/\beta)$. The condition that
its leading part vanish at a fixed common massless locus is
\begin{equation}
 \boldsymbol{\cF}^{\mathrm{in}}(\boldsymbol{x}^s)=0,
 \qquad \cF^{\mathrm{in}}_j(\boldsymbol{x})
 :=R_1F^{\mathrm{in}}_j(\boldsymbol{x}/R_1).
 \label{eq:IS_in_R3S1}
\end{equation}
If all retained chirals are massless there, this equals the
continued condition $\boldsymbol\cF(\boldsymbol{x}^s)=0$.
This equality follows directly from the threshold matching
in Eq.~\eqref{eq:thresholdmatching}.

Equation~\eqref{eq:IS_in_R3S1} suggests that the contribution of inner patches with $F_j^{\mathrm{in}}(\boldsymbol x^s/R_1)\neq0$ to the partition function is suppressed. Since on $S^3\times S^1$ Higgs sectors are lifted by curvature couplings, this equation can be interpreted as the absence of Coulomb-slice vacua in the presence of local FI terms, compatibly with \eqref{eq:wallvacua}.

\subsection{Progenitor: perturbative $\mathbb{R}^3$ potential from $S^3$}\label{app:R3_from_S3}

In this subsection we show that the above connection between $S^3\times S^1$ and $\mathbb{R}^3\times S^1$ has a simpler progenitor in a connection between $S^3$ and $\mathbb{R}^3$.

We start from $Z_{S^3}$ with $R_3$ reinstated:
    \begin{equation}
\begin{split}
    Z(\boldsymbol{m}_f)&=R^{r_G}_3\int_{-\infty}^\infty \frac{\mathrm{d}^{r_G}\sigma}{|W|}\ e^{-2\pi iR_{3}^2\,( k_{ij} \frac{\sigma_i\sigma_j}{2} \,+\,
    \zeta_j\sigma_j)\,+\,2\pi \, k^{}_{jR} \sigma_j R_{3}}\\
    &\hspace{3cm}\times\frac{\prod_\chi\prod_{\rho}\Gamma_h\big(i r_\chi\, +\rho\cdot {\boldsymbol{\sigma}}R_3+q^\chi\cdot \boldsymbol{m}_f R_3\big)}{\prod_{\alpha_+}\Gamma_h\big(\alpha_+\cdot{\boldsymbol{\sigma}}R_3\big)\Gamma_h\big(-\alpha_+\cdot{\boldsymbol{\sigma}}R_3\big)}\,,
    \end{split}\label{eq:S3_Z_gen}
\end{equation}
For simplicity, we have suppressed the contributions from $k_{RR},\,k_{\text{grav}},\,k_{fR},$ and $k_{ff}$ \cite{Closset:2019hyt}.

FI couplings may be included as masses for background topological symmetries: $\zeta_j=\sum_\ell {m_f}^\ell k_{j\ell}$, where $\ell$ runs over $U(1)$ flavor symmetries, including the $U(1)_J$ symmetries for which $k^{}_{jJ}$ is fixed to $1.$

The asymptotic determinant expansion and a perturbative
flat-space interpretation impose different conditions. Define
a cutoff
\begin{equation}
    \Lambda_3\qquad\text{subject to $R_3\Lambda_3\gg1$.}
\end{equation}
For a perturbative flat-space interpretation also require $e_0^2\ll\Lambda_3$, with fixed charge factors. We distinguish between
    \begin{equation}
    \begin{split}
        \text{the \textbf{\emph{far zone}} where}\quad &|\rho\cdot {\boldsymbol{\sigma}}+q^\chi\cdot\boldsymbol{m}_f|\ge\Lambda_3\quad\text{for all $\rho_\chi\neq0$},\\
        &\text{and}\quad |\alpha_+\cdot {\boldsymbol{\sigma}}|\ge\Lambda_3\quad\text{for all $\alpha_+$},
        \end{split}
    \end{equation}
and the rest of the Coulomb branch, which we decompose into
\begin{equation}
    \begin{split}
\text{\textbf{\emph{near zones}} where}\quad &|\rho\cdot {\boldsymbol{\sigma}}+q^\chi \cdot\boldsymbol{m}_f|<\Lambda_3\quad\text{for some $\rho_\chi\neq0$},\\
        &\text{or}\quad|\alpha_+\cdot {\boldsymbol{\sigma}}|<\Lambda_3\quad\text{for some $\alpha_+$}.
    \end{split}
\end{equation}
In short, near zones are where there are light charged multiplets, with \emph{light} meaning absolute real mass less than $\Lambda_3$. Note that near zones are around special points $\boldsymbol{\sigma}^p$ or $\boldsymbol{\sigma}^g$ satisfying
\begin{equation}
\begin{split}
         \rho\cdot {\boldsymbol{\sigma}^p}+q^\chi \cdot\boldsymbol{m}_f=0\quad&\text{for some $\rho_\chi\neq0$},\\
         \alpha_+\cdot {\boldsymbol{\sigma}^g}=0\quad&\text{for some $\alpha_+$}.
         \end{split}
    \end{equation}
We denote such $\boldsymbol{\sigma}^p,\boldsymbol{\sigma}^g$ collectively by $\boldsymbol{\sigma}^s.$

\paragraph{Note on inner/outer versus near/far.} If the 3d theory in question arises on the
inner patch of the Coulomb branch of a circle-compactified 4d theory, then the near
zone corresponds to an $O(\beta)$ neighborhood of $\boldsymbol x_s$, while the rest of the $O(\epsilon)$-sized inner
patch would count as the far zone. In this far-zone part of the inner patch, we have
charged multiplets that are heavy compared to the cutoff $\Lambda_3$, but are light compared
to the cutoff $\Lambda_M$. This could cause confusion if we had to simultaneously discuss
inner patches and near zones (as in \cite{ArabiArdehali:2025bub}, where the notions of intermediate zones and
middleweight multiplets were introduced to avoid confusion), but since in this work we
reserve the near/far discussions for the present subsection only, confusion regarding what is light and what is heavy is unlikely.

\subsection*{Far zone}

For $|\rho\cdot\boldsymbol\sigma+q^\chi\cdot\boldsymbol m_f|R_3\gg1$ and $|\alpha_+\cdot\boldsymbol\sigma|R_3\gg1$, inside the integrand we use the simplification
\begin{equation}
\frac{\prod_\chi \prod_{\rho
\in\Delta_\chi}\Gamma_h(i\,r_\chi +\rho\cdot
\boldsymbol{\sigma}R_3+q^\chi \cdot\boldsymbol{m}_f R_3)}{\prod_{\alpha_+}\Gamma_h(\pm\alpha_+\cdot
\boldsymbol{\sigma}R_3)}\approx e^{2\pi
i R_3^2\,Q^{(3)}(\boldsymbol{\sigma})-2\pi R_3
\,L^{(3)}(\boldsymbol{\sigma})},\label{eq:LagZ3dIntegrandAsy}
\end{equation}
valid up to a piecewise constant overall factor \cite{ArabiArdehali:2015ybk,Ardehali:2021irq}. The piecewise quadratic (resp. linear) function $Q^{(3)}(\boldsymbol{\sigma})$ (resp. $L^{(3)}(\boldsymbol{\sigma})$) is defined as\footnote{These functions were denoted $\tilde{Q}_{S^3},\tilde{L}_{S^3}$ in \cite{ArabiArdehali:2015ybk}.}
\begin{equation}
    \begin{split}
        Q^{(3)}(\boldsymbol{\sigma})&=-\frac{1}{4}\sum_{\chi}\sum_{\rho
\in\Delta_\chi}(\rho\cdot
\boldsymbol{\sigma}+q^\chi \cdot\boldsymbol{m}_f)|\rho\cdot
\boldsymbol{\sigma}+q^\chi \cdot\boldsymbol{m}_f|,\\
L^{(3)}(\boldsymbol{\sigma})&=-\frac{1}{2}\sum_{\chi}(r_\chi-1)\sum_{\rho
\in\Delta_\chi}|\rho\cdot
\boldsymbol{\sigma}+q^\chi \cdot\boldsymbol{m}_f|-\sum_{\alpha_+}|\alpha_+\cdot
\boldsymbol{\sigma}|.
    \end{split}
\end{equation}

The far zone consists of multiple disjoint connected components. On each connected component, the functions $Q^{(3)},L^{(3)}$ are analytic, and admit Taylor expansion. Denoting the $\boldsymbol{\sigma}$-independent terms by const, we get from the expansions
\begin{equation}
    \begin{split}
        Q^{(3)}(\boldsymbol{\sigma})&=-k^\text{1-loop}_{ij}\frac{\sigma_i\sigma_j}{2}-\zeta_j^\text{1-loop}\sigma_j+\text{const}\,,\\
        L^{(3)}(\boldsymbol{\sigma})&=-k^\text{1-loop}_{jR}\sigma_j+\text{const}\,.
    \end{split}
\end{equation}
Here $k_{ij}^\text{1-loop},\zeta_{j}^\text{1-loop},k_{jR}^\text{1-loop}$ are the one-loop corrections generated by integrating out the charged multiplets which, by the definition of the far zone, are all heavy there. We thus have (up to nonzero overall factors constant on each connected component) %
\begin{equation}
\begin{split}
    Z_\text{far zone}(\boldsymbol{m}_f)&\xrightarrow{R_3\to\infty}R^{r_G}_3\int_\text{far zone} \!\!{\mathrm{d}^{r_G}\sigma}
    \ \ e^{-2\pi iR_{3}^2\,( k^\text{eff}_{ij} \frac{\sigma_i\sigma_j}{2} \,+\,
    \zeta^\text{eff}_j\sigma_j)\,+\,2\pi \, k^\text{eff}_{jR} \sigma_j R_{3}},%
    \end{split}\label{eq:S3_Z_far_simp}
\end{equation}
with the effective couplings obtained as the sum of the classical and one-loop pieces.

If similarly to \eqref{eq:AZ} we consider
\begin{equation}
    \boldsymbol{m}_f R_3= \boldsymbol{m}^0_f R_3+\hat{\boldsymbol{m}}_f R_3,\label{eq:m_inf_fin}
\end{equation}
and keep $\hat{\boldsymbol{m}}_f R_3$ constant while sending $\boldsymbol{m}^0_f R_3\to\infty,$ we get corrections of the form $k_{j\ell}\,{\hat{m}_f}^{\ \ell}$ which we package into $\zeta_j^\text{eff}$ analogously to \eqref{eq:xi_plus_kjfmf}.

The leading $R_3^2$ phase has stationary equation
\begin{equation}
 0=k^{\mathrm{eff}}_{ij}\sigma_i+\zeta_j^{\mathrm{eff}}
 =k_{ij}\sigma_i+\zeta_j-\partial_{\sigma_j}Q^{(3)}
 =F_j(\boldsymbol\sigma).
 \label{eq:S3stationarity}
\end{equation}
With a positive nonsingular kinetic matrix, which we expect in the far zone as in Section~\ref{subsubsec:controlledCH}, this is the
perturbative D-flat condition
$V_{D,\mathrm{quad}}\propto F_i(\mathcal G^{-1})_{ij}F_j=0$.

Extended saddles on the far zone arise along some direction $\hat{\boldsymbol{g}}$ when the second derivative of the coefficient of $R_3^2$ in \eqref{eq:S3_Z_far_simp} satisfies:
\begin{equation}
    k^\text{eff}_{ij}\hat{g}_i=0,\qquad\text{for all $j$}.
\end{equation}
For an allowed magnetic charge in this kernel, the induced gauge charge is zero since $c_j(\boldsymbol{m})=-k^\text{eff}_{ij}m_i.$

We thus see that the first derivative $\partial_{{\sigma}_j}$ of $Q^{(3)}$ encodes the 1-loop correction to $F_j(\boldsymbol{\sigma})$, while the second derivative $\partial_{{\sigma}_i}\partial_{{\sigma}_j}$ of $Q^{(3)}$ encodes the 1-loop correction to $k_{ij}$ (and hence the gauge charge of BPS monopoles).

The leading $R_3^2$ phase therefore reproduces the perturbative Coulomb-slice zero equations in the controlled far zone.

\subsection*{Near zones}

Choose a reference point $\boldsymbol\sigma^s$ and set
$\tilde{\boldsymbol\sigma}=R_3(\boldsymbol\sigma-\boldsymbol\sigma^s)$.
Let $F_j^{\mathrm{nz}}$ and $k_{ij}^{\mathrm{nz}}$ be the coefficients
after integrating out only the fields heavy in this zone, and
$\mu_\rho^s=R_3(\rho\cdot\boldsymbol\sigma^s+
q^\chi\cdot\boldsymbol m_f)$. The local integral is
\begin{equation}
\begin{split}
 Z_{\mathrm{nz}}&\sim
 \int_{R_3(\mathrm{nz}-\boldsymbol\sigma^s)}d^{r_G}\tilde\sigma\,
 e^{-2\pi i[\frac12k^{\mathrm{nz}}_{ij}\tilde\sigma_i\tilde\sigma_j
 +R_3F_j^{\mathrm{nz}}(\boldsymbol\sigma^s)\tilde\sigma_j]
 +2\pi k^{\mathrm{nz}}_{jR}\tilde\sigma_j}\\
 &\quad\times
 \frac{\prod_{\rho\in L_{\chi,\mathrm{nz}}}
 \Gamma_h(i r_\chi+\rho\cdot\tilde{\boldsymbol\sigma}+\mu_\rho^s)}
 {\prod_{\alpha\in L_{V,\mathrm{nz}}}
 \Gamma_h(\alpha\cdot\tilde{\boldsymbol\sigma}+\mu_\alpha^s)}.
\end{split}
\label{eq:S3_Z_near}
\end{equation}
The original measure factor $R_3^{r_G}$ cancels the Jacobian;
locally constant factors and the Weyl normalization are suppressed.
For finite scaled masses and bounded $\tilde{\boldsymbol\sigma}$,
the absence of a growing linear phase requires
\begin{equation}
 F_j^{\mathrm{nz}}(\boldsymbol\sigma^s)=0
 \label{eq:IS_in}
\end{equation}
at leading order in the large-radius limit. This suggests that the contribution of the near zones with $F_j^{\mathrm{nz}}(\boldsymbol\sigma^s)\neq0$ to the partition function is suppressed.

\section{Compatibility with $\mathbb{R}^2\times T^2$}
\label{app:QfromW}

For vanishing flavor holonomies, the FI-deformed combination
$Q-\tilde{\boldsymbol{\zeta}}^{(4)}\cdot\boldsymbol{x}$
can also be obtained from the large-$r_1$ limit of
$\mathbb{R}^2\times S^1_{r_1}\times S^1_{r_2}$.
Here $r_2$ is the radius denoted $R_1$ in the main text, so
the dimensionless renormalized FI parameters are
\begin{equation}
 \tilde{\boldsymbol{\zeta}}^{(4)}
 =(2\pi r_2)^2\boldsymbol{\zeta}^{(4)},
\end{equation}
with nonzero components allowed only along abelian factors
of the four-dimensional gauge group. The exact compactified vacuum equations are Bethe equations,
\begin{equation}
    \exp(2\pi i\partial_{u_j}W)=1\,,
\end{equation}
with logarithmic branches
and global identifications \cite{Closset:2017bse},
in terms of the 2d $\mathcal{N}=(2,2)$ twisted superpotential of the effective theory on $\mathbb{R}^2$. We show in this appendix that on a fixed logarithmic branch away from thresholds, with bounded
branch integers
\begin{equation}
 W\xrightarrow{r_1\to\infty}
 -i\frac{r_1}{r_2}
 \left(
 Q(\boldsymbol{x})
 -\tilde{\boldsymbol{\zeta}}^{(4)}\cdot\boldsymbol{x}
 \right),
 \label{eq:W_vs_Q}
\end{equation}
at fixed $r_2$ and $\boldsymbol{\zeta}^{(4)}$.
The Bethe equations then imply
$\partial_{x_j}Q=\tilde{\zeta}^{(4)}_j$
at leading order, reproducing the Coulomb-slice equations
$F_j=0$.
(We emphasize that this does not quite establish a one-to-one correspondence of all
Bethe vacua with stationary points of
$Q-\tilde{\boldsymbol{\zeta}}^{(4)}\cdot\boldsymbol{x}$;
branch integers scaling with the limit and wall sectors
require separate analysis.)

The contribution of a 4d $\mathcal{N}=1$ chiral multiplet
with gauge weight $\rho$ to the effective twisted
superpotential is \cite{Closset:2017bse}\footnote{We have
removed a term $-(\rho\cdot\boldsymbol{u})^3/(6\tau)$
from the RHS. Its sum over all weights vanishes by
cancellation of the cubic gauge anomaly, so this change
has no effect in the full anomaly-free theory at vanishing
flavor holonomies. For an individual chiral multiplet,
however, \eqref{eq:W_2d} uses a different regularization
of the infinite KK sums from \cite{Closset:2017bse}.
Consequently, its modular transformation is
\begin{equation}
 W_\rho\left(\frac{\boldsymbol{u}}{\tau};
 -\frac{1}{\tau}\right)
 =\frac{W_\rho(\boldsymbol{u};\tau)}{\tau}
 -\frac{(\rho\cdot\boldsymbol{u})^3}{6\tau^2}
 +\frac{\rho\cdot\boldsymbol{u}}{4\tau},
 \label{eq:W_modular}
\end{equation}
with the opposite sign for the cubic term compared with
Eq.~(2.33) of \cite{Closset:2017bse}. This transformation
concerns the chiral contribution, before adding the
classical FI term.\label{fn:scheme}}
\begin{equation}
\begin{split}
 W_\rho(\boldsymbol{u};\tau)
 &=\frac{(\rho\cdot\boldsymbol{u})^2}{4}
 -\frac{(\rho\cdot\boldsymbol{u})\tau}{12}
 +\frac{1}{24}\\
 &\quad+\frac{1}{(2\pi i)^2}
 \sum_{k=0}^{\infty}
 \left[
 \operatorname{Li}_2
 \left(e^{2\pi i(\rho\cdot\boldsymbol{u}+k\tau)}\right)
 -
 \operatorname{Li}_2
 \left(e^{2\pi i(-\rho\cdot\boldsymbol{u}+(k+1)\tau)}\right)
 \right].
\end{split}
\label{eq:W_2d}
\end{equation}
Here
$\tau=ir_2/r_1$ and
$\boldsymbol{u}=\boldsymbol{x}+\tau\boldsymbol{y}$
is the complex gauge holonomy.
In an integral basis for the holonomy lattice, the
components of $\boldsymbol{x}$ and $\boldsymbol{y}$
are 1-periodic.

The four-dimensional FI term reduces to a classical
linear twisted superpotential.
In our normalization, and with no additional
two-dimensional theta term, it is
\begin{equation}
\begin{split}
 W_{\mathrm{FI}}(\boldsymbol{u};\tau)
 &=-\frac{
 \tilde{\boldsymbol{\zeta}}^{(4)}\cdot\boldsymbol{u}
 }{\tau}\\
 &=i(2\pi r_1)(2\pi r_2)\,
 \boldsymbol{\zeta}^{(4)}\cdot\boldsymbol{u}.
\end{split}
\label{eq:W_FI}
\end{equation}
Thus its coefficient is proportional to the torus area,
as required by dimensional reduction.
The full twisted superpotential is
\begin{equation}
 W(\boldsymbol{u};\tau)
 =W_{\mathrm{FI}}(\boldsymbol{u};\tau)
 +\sum_\rho W_\rho(\boldsymbol{u};\tau),
 \label{eq:W_total}
\end{equation}
where the sum includes all chiral multiplets and
multiplicities.
The linear FI term is written on a local lift of the
holonomy torus; its shifts under large gauge
transformations are independent of $\boldsymbol{u}$.

Take $\tau\to i0^+$ with
$\boldsymbol{u}=\boldsymbol{x}+\tau\boldsymbol{y}$,
bounded $\boldsymbol{y}$, and every charged argument
$\rho\cdot\boldsymbol{x}$ a fixed positive distance
from $\mathbb{Z}$.
Replacing $\boldsymbol{u}$ by $\boldsymbol{x}$ changes
only the $O(1)$ term.
For a real scalar argument $v\notin\mathbb{Z}$, write
\begin{equation}
 S(v;\tau):=\sum_{k=0}^{\infty}
 \left[
 \operatorname{Li}_2(e^{2\pi iv+2\pi ik\tau})
 -
 \operatorname{Li}_2(e^{-2\pi iv+2\pi i(k+1)\tau})
 \right].
\end{equation}
Euler--Maclaurin summation gives the leading term
\begin{equation}
\begin{split}
 S(v;\tau)
 &=\frac{1}{2\pi i\tau}
 \int_1^0\frac{dz}{z}
 \left[
 \operatorname{Li}_2(e^{2\pi iv}z)
 -
 \operatorname{Li}_2(e^{-2\pi iv}z)
 \right]+O(1)\\
 &=-\frac{
 \operatorname{Li}_3(e^{2\pi iv})
 -
 \operatorname{Li}_3(e^{-2\pi iv})
 }{2\pi i\tau}+O(1)\\
 &=\frac{(2\pi i)^2}{6\tau}\,
 \overline B_3(v)+O(1).
\end{split}
\label{eq:Li3_simp}
\end{equation}
The identity on the last line also underlies the
Cardy-like asymptotics of the 4d index derived in
\cite{Choi:2018hmj}.

Combining \eqref{eq:Li3_simp} with \eqref{eq:W_2d}
gives
\begin{equation}
 W_\rho(\boldsymbol{u};\tau)
 =\frac{\overline B_3(\rho\cdot\boldsymbol{x})}{6\tau}
 +O(1)
 =\frac{Q_\rho(\boldsymbol{x})}{\tau}+O(1).
\end{equation}
Meanwhile, the classical FI contribution satisfies
\begin{equation}
 W_{\mathrm{FI}}(\boldsymbol{u};\tau)
 =-\frac{
 \tilde{\boldsymbol{\zeta}}^{(4)}\cdot\boldsymbol{x}
 }{\tau}
 -\tilde{\boldsymbol{\zeta}}^{(4)}\cdot\boldsymbol{y}.
\end{equation}
Summing over all weights therefore yields
\begin{equation}
\begin{split}
 W(\boldsymbol{u};\tau)
 &=\frac{1}{\tau}
 \left[
 Q(\boldsymbol{x})
 -\tilde{\boldsymbol{\zeta}}^{(4)}\cdot\boldsymbol{x}
 \right]+O(1)\\
 &=-i\frac{r_1}{r_2}
 \left[
 Q(\boldsymbol{x})
 -\tilde{\boldsymbol{\zeta}}^{(4)}\cdot\boldsymbol{x}
 \right]+O(1),
\end{split}
\end{equation}
which establishes \eqref{eq:W_vs_Q}.
The remainder is bounded on compact subsets away from
the charged thresholds, for bounded $\boldsymbol{y}$
and fixed FI inputs.

To include flavor holonomies, write
$\boldsymbol{\nu}=\boldsymbol{\xi}+\tau\boldsymbol{\eta}$
and set
\begin{equation}
 z_\rho=\rho\cdot\boldsymbol{u}
             +q^\chi\cdot\boldsymbol{\nu},
 \qquad
 v_\rho=\rho\cdot\boldsymbol{x}+q^\chi\cdot\boldsymbol{\xi}.
\end{equation}
We hold $\boldsymbol{\xi}$ fixed, keep
$\boldsymbol{y}$ and $\boldsymbol{\eta}$ bounded, and
stay a fixed positive distance from the thresholds
$v_\rho\in\mathbb Z$.
Extending the subtraction convention of \eqref{eq:W_2d} discussed in Footnote~\ref{fn:scheme} by replacing $\rho\cdot\boldsymbol{u}$ with $z_\rho$,
the calculation in \eqref{eq:Li3_simp} gives
\begin{equation}
W_\rho(\boldsymbol{u},\boldsymbol{\nu};\tau)
 =\frac{\overline B_3(v_\rho)}{6\tau}+O(1).
\end{equation}
Consequently\footnote{Let $W_{\mathrm{ref}}$ denote the twisted superpotential
in the regularization scheme of \cite{Closset:2017bse}, including
the same classical FI term. The difference is
\begin{equation}
\begin{split}
 \Delta W:=W-W_{\mathrm{ref}}
 =\frac{1}{6\tau}\sum_\rho z_\rho^3=\frac{\mathcal A_{ij\alpha}}{2\tau}
       u_i u_j\nu_\alpha
  +\frac{\mathcal A_{i\alpha\beta}}{2\tau}u_i\nu_\alpha\nu_\beta
  +\frac{\mathcal A_{\alpha\beta\gamma}}{6\tau}\nu_\alpha\nu_\beta\nu_\gamma ,
\end{split}
\end{equation}
where repeated indices are summed and $\mathcal A_{ij\alpha}
=\sum_\rho\rho_i\rho_jq^\chi_\alpha,\,
 \mathcal A_{i\alpha\beta}
=\sum_\rho\rho_iq^\chi_\alpha q^\chi_\beta,\,
 \mathcal A_{\alpha\beta\gamma}
=\sum_\rho q^\chi_\alpha q^\chi_\beta q^\chi_\gamma.$
Here only the pure gauge cubic was eliminated using
$\sum_\rho\rho_i\rho_j\rho_k=0$.
In particular, the mixed terms contribute
$\partial_{u_i}\Delta W
 =\frac{1}{\tau}\left[
   \mathcal A_{ij\alpha}u_j\nu_\alpha
   +\frac12\mathcal A_{i\alpha\beta}
       \nu_\alpha\nu_\beta\right].$
Thus, in the scheme of \cite{Closset:2017bse}, the leading Bethe
equations read
\begin{equation}
\partial_{x_i}Q(\boldsymbol{x};\boldsymbol{\xi})
 =\tilde\zeta_i^{(4)}
  +\mathcal A_{ij\alpha}x_j\xi_\alpha
  +\frac12\mathcal A_{i\alpha\beta}\xi_\alpha\xi_\beta.
\end{equation}
For flavor symmetries satisfying the mixed-anomaly
conditions
$\mathcal A_{ij\alpha}=\mathcal A_{i\alpha\beta}=0$
of \cite[Eq.~(2.22)]{Closset:2017bse}, the schemes differ by the background term
$\Delta W=\mathcal A_{\alpha\beta\gamma}
\nu_\alpha\nu_\beta\nu_\gamma/(6\tau)$, independent of the gauge holonomies, leaving the exact gauge Bethe equations unchanged.
Under the same assumptions as above,
their leading limit is
$\partial_{x_i}Q(\boldsymbol{x};\boldsymbol{\xi})
=\tilde\zeta_i^{(4)}$, reproducing $F_i=0$ without
requiring $A_{\alpha\beta\gamma}$ to vanish.}
\begin{equation}
\begin{split}
 Q(\boldsymbol{x};\boldsymbol{\xi})
 &=\frac16\sum_\rho
   \overline B_3(\rho\cdot\boldsymbol{x}
                   +q^\chi\cdot\boldsymbol{\xi}),\\
 W(\boldsymbol{u},\boldsymbol{\nu};\tau)
 &=\frac{1}{\tau}\left[
   Q(\boldsymbol{x};\boldsymbol{\xi})
   -\tilde{\boldsymbol{\zeta}}^{(4)}
       \cdot\boldsymbol{x}\right]+O(1).
\end{split}
\end{equation}

\section{On zeta-regularization of KK sums}
\label{app:KKren}

The zeta-regularized expressions
\eqref{eq:Fj} and \eqref{eq:kout} admit a direct interpretation
in terms of cutoff sums and local subtractions. The induced
gauge--gauge CS coefficient is a finite threshold coefficient,
whereas the D-linear coefficient can contain a power-divergent
FI contribution. See
\cite{Poppitz:2008hr,Corvilain:2017luj} for earlier discussions of KK summed CS couplings.

For a chiral KK mode, write $M_{n,\rho}=\frac{n+u_\rho}{R_1},$ $u_\rho=\rho\cdot\boldsymbol{x}+q^\chi\cdot\boldsymbol{\xi}.$ In our conventions, its threshold contributions are \cite{Intriligator:2013lca}
\begin{equation}
 \delta F_i=\frac12\rho_i|M_{n,\rho}|,
 \qquad
 \delta k_{ij}
 =\frac12\rho_i\rho_j\operatorname{sgn}(M_{n,\rho}).
 \label{eq:KKren_single}
\end{equation}
The CS formulas below apply away from massless thresholds;
modes becoming light must instead be retained explicitly.

\paragraph{An explicit cutoff and the FI subtraction.}

Introduce a dimensionless regulator
$t=(\Lambda_{\mathrm{UV}}R_1)^{-1}>0$, where
$\Lambda_{\mathrm{UV}}$ is distinct from (and assumed much larger than) the matching cutoff
$\Lambda_M$. For $v=\{u\}$, geometric summation gives
\begin{align}
 T_t(u)
 &:=\sum_{n\in\mathbb Z}|n+u|e^{-t|n+u|}
 \nonumber\\
 &=-\partial_t
 \frac{e^{-tv}+e^{-t(1-v)}}{1-e^{-t}}
 =\frac{2}{t^2}-\overline B_2(u)+O(t^2).
 \label{eq:KKren_abs}
\end{align}
Thus the Bernoulli polynomial arises as the finite part of an
ordinary convergent sum, without analytic continuation.

Using \eqref{eq:KKren_abs}, the regulated D-linear coefficient is
\begin{equation}
 F_i^{\Lambda_{\mathrm{UV}}}
 =
 \zeta_i^{\mathrm{bare}}(\Lambda_{\mathrm{UV}})
 +\Lambda_{\mathrm{UV}}^2R_1\sum_\rho\rho_i
 -\frac{1}{2R_1}\sum_\rho\rho_i\overline B_2(u_\rho)
 +o(1).
 \label{eq:KKren_Fbare}
\end{equation}
The divergent term is independent of the holonomy and multiplies
the local FI operator. The subtraction
\begin{equation}
 \zeta_i^{\mathrm{bare}}(\Lambda_{\mathrm{UV}})
 =
 \zeta_i^{\mathrm{ren}}
 -\Lambda_{\mathrm{UV}}^2R_1\sum_\rho\rho_i,
 \qquad
 \zeta_i^{\mathrm{ren}}
 =\frac{\tilde\zeta_i^{(4)}}{R_1},
 \label{eq:KKren_FIct}
\end{equation}
therefore reproduces \eqref{eq:Fj}. The coefficient of the
power divergence is regulator dependent; the displayed value
belongs to the chosen cutoff prescription.
For a semisimple gauge generator,
$\sum_{\rho\in\mathcal R}\rho_i
=\operatorname{Tr}_{\mathcal R}T_i=0$
in each representation. For an abelian gauge factor, the same
trace vanishes when the mixed gauge--gravitational anomaly
cancels. In these cases the divergent FI contribution cancels
in the complete spectrum.

\paragraph{The finite CS coefficient.}

The corresponding regulated sign sum, for $u\notin\mathbb Z$, is
\begin{align}
 S_t(u)
 &:=\sum_{n\in\mathbb Z}
 \operatorname{sgn}(n+u)e^{-t|n+u|}
 \nonumber\\
 &=\frac{e^{-t\{u\}}-e^{-t(1-\{u\})}}{1-e^{-t}}
 =1-2\{u\}+O(t^2).
 \label{eq:KKren_sign}
\end{align}
There is no divergent CS subtraction in this prescription.
Its finite limit gives
\begin{equation}
 k_{ij}^{\mathrm{ren}}
 =
 \sum_\rho\rho_i\rho_j
 \left(\frac12-\{u_\rho\}\right)
 =
 -\sum_\rho\rho_i\rho_j\overline B_1(u_\rho)
 =
 \frac{\partial F_i^{\mathrm{ren}}}{\partial\sigma_j},
 \label{eq:KKren_k}
\end{equation}
in agreement with \eqref{eq:kout}. Gauge-anomaly cancellation
makes this coefficient constant within each chamber.

A different elementary prescription cuts off the KK labels
symmetrically. For sufficiently large $N$,
\begin{equation}
 \sum_{n=-N}^{N}\operatorname{sgn}(n+u)
 =1+2\lfloor u\rfloor.
 \label{eq:KKren_label}
\end{equation}
Comparing this with \eqref{eq:KKren_sign} gives
\begin{equation}
 k_{ij}^{\mathrm{spectral}}-k_{ij}^{\mathrm{label}}
 =
 -\sum_\rho\rho_i\rho_j u_\rho
 =
 -\mathcal A_{ijk}x_k
 -\mathcal A_{ij\alpha}\xi_\alpha,
 \label{eq:KKren_anomaly}
\end{equation}
where
\begin{equation}
 \mathcal A_{ijk}
 =\sum_\rho\rho_i\rho_j\rho_k,
 \qquad
 \mathcal A_{ij\alpha}
 =\sum_\rho\rho_i\rho_j q^\chi_\alpha.
\end{equation}
The two prescriptions therefore agree when the pure gauge
anomaly and the relevant mixed gauge--gauge--flavor anomaly
vanish.

The polynomial differences in \eqref{eq:KKren_anomaly} are not
gauge-invariant counterterms; they encode
anomalous Ward identities. Indeed, a term
$K(\sigma)A\wedge dA$ has a gauge variation proportional to
$dK$. Explicit Pauli--Villars calculations likewise exhibit
anomaly-dependent differences between regulator prescriptions
\cite[Secs.~2.3--2.5]{Corvilain:2017luj}.

\bibliographystyle{JHEP}
\bibliography{biblio}

\end{document}